\documentclass[%
 reprint,
 amsmath,amssymb,
 aps,
]{revtex4-2}

\usepackage{graphicx}
\usepackage{dcolumn}
\usepackage{bm}
\usepackage{tikz}

\definecolor{colEta1}{rgb}{0, 0.7, 0.6}
\definecolor{colEta2}{rgb}{0, 0, 1}
\definecolor{colEta3}{rgb}{1, 0.5, 0}
\definecolor{colEta4}{rgb}{0.8, 0.4, 1}
\definecolor{colEta5}{rgb}{0, 1, 1}
\definecolor{colEta6}{rgb}{1, 0, 0}
\definecolor{colEta7}{rgb}{0.6, 0.4, 0.4}
\definecolor{colEta8}{rgb}{1, 0.7, 0.2}
\definecolor{colAlpha}{rgb}{0.2, 0.7, 1}
\definecolor{colBeta}{rgb}{0.7, 0.4, 0.7}

\begin{document}

\preprint{APS/123-QED}

\title{The Veneziano Amplitude in any Dimension
\\
and a Virasoro-Shapiro Partial Amplitude}

\author{Christian Baadsgaard Jepsen}%
\affiliation{School of Physics, Korea Institute for Advanced Study, Seoul 02455, Korea
}
\date{\today}

\begin{abstract}
This paper demonstrates how the Veneziano partial amplitude of bosonic string theory admits a generalization to world-(hyper)surfaces of any dimension $d$. In particular, for $d=2$, by carving up the worldsheet integral according to stipulations imposed on conformal cross ratios, the Virasoro-Shapiro full amplitude can be decomposed into a sum of three partial amplitudes. The amplitudes obtained on generalizing the Veneziano amplitude all contain tachyons. To explore candidate tachyon-free and supersymmetric versions of these amplitudes, a new bootstrap principle is introduced and applied, which demands that towers of residues alternate between all-even and all-odd spin partial waves.
\end{abstract}

\maketitle


\section{\label{Sec1}Introduction and Summary}

Is it possible to distinguish between two fundamental strings that are both in the ground state? In the case of the open bosonic string the answer is yes. The ground state is a scalar and is unique but the string endpoints provide a means of telling strings apart, as these may be situated on different branes and can carry charge. At the level of scattering amplitudes, this distinguishability expresses itself in the existence of partial amplitudes, into which full amplitudes may be decomposed. In terms of Mandelstam invariant $s = -(k_1+k_2)^2$, $t = -(k_1+k_4)^2$, and $u = -(k_1+k_3)^2$, where a factor of $\sqrt{\alpha'}$ has been absorbed into the momenta, the full Veneziano amplitude \cite{Veneziano:1968yb},
\begin{align}
&
A^{(1)}(s,t) = \pi^{1/2}\frac{\Gamma(\frac{-s-1}{2})\Gamma(\frac{-t-1}{2})\Gamma(\frac{-u-1}{2})}{\Gamma(\frac{s+2}{2})\Gamma(\frac{t+2}{2})\Gamma(\frac{u+2}{2})}\,,
\end{align}
with $s+t+u=-4$, is a function totally symmetric in $s$, $t$, and $u$, in keeping with the amplitude describing the scattering of four identical tachyons. Meanwhile, the Veneziano partial amplitude,
\begin{align}
A_p^{(1)}(s,t) = \frac{\Gamma(-s-1)\Gamma(-t-1)}{\Gamma(-s-t-2)}\,,
\end{align}
exhibits crossing symmetry only under the interchange of $s$ and $t$, so that amplitude may describe scattering of the type $AB \rightarrow AB$ of four objects not all identical. In computing the total amplitude of such a process one sums over partial amplitudes in the different scattering channels, but the partial amplitudes may be dressed in Chan-Paton factors \cite{Paton:1969je} associated to the string endpoints. Only in the case when all Chan-Paton factors are equal does the weighted sum of partial amplitudes recover the full $s,t,u$-symmetric Veneziano amplitude:
\begin{align}
\label{A1sum}
A^{(1)}(s,t) = A_p^{(1)}(s,t)+A_p^{(1)}(t,u)+A_p^{(1)}(u,s)\,.
\end{align}
Knowledge of the partial amplitude is thus strictly superior to knowledge of the full amplitude. Moreover, information on the spectrum of the underlying theory can be gleaned from the poles of the amplitude, which occur as virtual particles go on-shell. From the poles of $A_p(s,t)$, situated at $s\in -1+\mathbb{N}_0$, one can read off the mass levels of the open bosonic string. But on taking the sum \eqref{A1sum}, exchanged states with odd spin cancel between the different scattering channels, leaving behind only the poles at $s\in -1+2\mathbb{N}_0$ so that the $s,t,u$-symmetric full amplitude contains data only on the spin-even spectrum of the theory. \footnote{To see the absence of odd-spin exchange beyond the present case of four external scalars, it can be useful to adopt the spinor-helicity formalism, see e.g. \cite{Arkani-Hamed:2017jhn}.}
\\[2mm]
Let us now turn to closed strings, which do not have endpoints that can be attached to branes and endowed with Chan-Paton factors. Is it possible to distinguish between two such strings in the ground state in the absence of endpoints where charge can reside? Up until now, the Virasoro-Shapiro closed string four-tachyon amplitude \cite{Virasoro:1969me,Shapiro:1970gy}, which is fully permutation symmetric in all four external momenta, has not been known to admit a decomposition into any kind of partial amplitudes with lowered permutation symmetry. But the existence of such a decomposition is a necessary mathematical prerequisite in order for distinguishable, unexcited closed strings to be able to scatter. And the present paper will show that this condition is in fact met---there does exist a partial amplitude for the closed bosonic string, given in terms of a ${}_3\widetilde{F}_2$ regularized hypergeometric function:
\begin{align}
\label{d2sumFormula}
&A_p^{(2)}(s,t)=
-\sum_{n=0}^\infty\frac{(-1)^n\pi\,\Gamma(\frac{-t-2}{2})}{n!\,\Gamma(\frac{n+2}{2})}\,\frac{1}{s-n+2}\,\times\\ \nonumber
&
{}_3\widetilde{F}_2\Big[\Big\{\hspace{-1mm}-n,\,-\frac{n}{2},\,-\frac{t+n+2}{2}\Big\};\Big\{\frac{2-n}{2},-\frac{t+2n+2}{2}\Big\};1\Big]
.
\end{align}
Barring the important caveat that the poles of $A_p^{(2)}(s,t)$ indicate that the spectrum now contains not one but two tachyons, $A_p^{(2)}(s,t)$ satisfies, to all appearances, every condition required by a physical scattering amplitude, and when summed over the three four-point scattering channels, it reproduces the Virasoro-Shapiro amplitude. 
\\[2mm]
Beyond strings, one may ask if the concept of partial amplitudes generalizes more broadly to other extended objects and can ultimately pave a way to describe open branes from an amplitude perspective. A simple setting in which to address this question can be found in the context of the Brower-Goddard dual models \cite{Brower:1971nd} obtained by uplifting the Koba-Nielsen \cite{Koba:1969rw} worldsheet integrals for the Veneziano and Virasoro-Shapiro amplitudes to higher-dimensional conformally symmetric integrals: 
\begin{align}
\label{intgeneral}
A^{(d)}(s,t)= \int_{(\mathbb{R}^d)^4}  d\Omega_4^{(d)}\,\prod_{i=1}^{3}\prod_{j=i+1}^4|\vec{x}_j-\vec{x}_i|^{2k_i\cdot k_j} \,,
\end{align}
with the integration measure given by
\begin{align}
& d\Omega_4^{(d)} \equiv\, 
|\vec{x}_a-\vec{x}^{\,0}_b|\,
|\vec{x}_b-\vec{x}^{\,0}_c|\,
|\vec{x}_c-\vec{x}^{\,0}_a|
\, \times
\\ \nonumber
&
\bigg[\prod_{i=1}^4 d^d\vec{x}_i\bigg]
\delta(\vec{x}_a-\vec{x}_a^{\,0})\,
\delta(\vec{x}_b-\vec{x}_b^{\,0})\,
\delta(\vec{x}_c-\vec{x}_c^{\,0})
\,,
\end{align}
where the subscripts $a$, $b$, $c$ are any three distinct indices from one to four \footnote{As written, the integral \eqref{intgeneral} is only independent of the choice of $\vec{x}_a^0$, $\vec{x}_b^0$, $\vec{x}_c^0$ when the external particles are tachyons with the specific mass stated in the main text. To retain the independence for other mass values, it is necessary to introduce additional powers of norms of differences to the integrand, as in the field known as $Z$-theory \cite{Carrasco:2016ldy}.}. The integrand for $A^{(d)}(s,t)$ possesses a $d$-dimensional conformal symmetry that renders its value independent of the choice of $\vec{x}^{\,0}_a$, $\vec{x}^{\,0}_b$, $\vec{x}^{\,0}_c$  provided that the external momenta are conserved and satisfy the tachyonic on-shell condition $(k_i)^2=d$, in which case the integral evaluates to the simple answer \cite{Brower:1971nd,Natsuume:1993ix}
\begin{align}
\label{Ad}
A^{(d)}(s,t)=\pi^{d/2}\frac{\Gamma(\frac{-d-s}{2})\Gamma(\frac{-d-t}{2})\Gamma(\frac{-d-u}{2})}{\Gamma(\frac{2d+s}{2})\Gamma(\frac{2d+t}{2})\Gamma(\frac{2d+u}{2})}\displaystyle\,.
\end{align}
As the Brower-Goddard amplitudes arise on integrating over higher-dimensional world-(hyper)surfaces, it seems their most plausibly interpretation is that they furnish a model for brane scattering, though theoretical attempts at studying such processes are known to be mired in obstacles of a fundamental nature \cite{deWit:1988xki,Green:1991pa}. In previous work, Emil Bjerrum-Bohr and the present author \cite{Bjerrum-Bohr:2024wyw} adopted a specific choice of moduli-fixing and determined analytically the odd-$d$ Brower-Goddard partial amplitudes. The present paper complements this work by providing the general prescription for obtaining partial amplitudes valid for any choice of moduli-fixing and by determining the even-$d$ partial amplitudes, thereby establishing the uplift of the Veneziano amplitude to any $d$.
\\[2mm]
It should be emphasized that $d$ here is defined as the dimensionality of the integral \eqref{intgeneral}. In the first two instances, the physical interpretation of this integral is known: for $d=1$ the integral reduces to the boundary of the worldsheet of the open string, and for $d=2$ the integral runs over all of the worldsheet of the closed string. Thus $d=1$ and $d=2$ both represent one-dimensional dynamical objects. Whether the individual cases of $d>2$ are best interpreted as integrals over worldvolumes of $(d-1)$-dimensional objects or over boundaries of worldvolumes of $d$-dimensional objects cannot be determined with certitude from the pure amplitude perspective of the present paper.
\\[2mm]
The $d>2$ partial amplitudes contain tachyons and massless higher-spin particles in their spectra. In the case of bosonic string theory, the tachyon problem is known to find its resolution in the superstring and the GSO projection \cite{Gliozzi:1976qd}. Until it can be ruled out that a similar cure exists in a theory accounting for the higher-$d$ amplitudes, they should perhaps not be dismissed outright. And indeed the final part of the present paper will seek argue that existing $S$-matrix bootstrap conditions and EFT-bounds are not sufficiently stringent to rule out families of higher-$d$ superamplitudes.
\\[2mm]
To hone in on the possible presence or absence of new families of superamplitudes, the present paper introduces a new string theory motivated bootstrap condition, which may be dubbed ``level-spin parity", and which requires the residues of an amplitude in a given scattering channel to alternate at increasing energy between containing only even-spin and only-odd spin partial waves. The condition greatly reduces the function space to consider but still leaves behind an infinite discrete set of functions of speculative physical interest, with the simplest of these functions, call it $A_s^{(2)}(s,t)$, being given by \footnote{The convention adopted here is to strip off from the tentative superamplitude a factor of $s^2$  arising from the supermomentum-conserving delta function $\delta^{8}(\mathcal{Q})$ so that the function considered, sometimes called $f(s,t)$ in the literature, exhibits crossing symmetry: $A_s^{(2)}(s,t)=A_s^{(2)}(t,s)$.}
\begin{align}
\label{As2formula}
&\hspace{10mm}A_s^{(2)}(s,t)= -\frac{\Gamma(-s-1)\Gamma(-t-1)}{\Gamma(2-s-t)}+
\\ \nonumber &
\hspace{3mm}\frac{1/6}{\prod_{n=-2}^1(s+n)(t+n)}
\bigg(
28-14(s+t)-4(s^2+t^2)+
\\  \nonumber &
2(s^3+t^3)\hspace{-0.3mm}-\hspace{-0.3mm}9st\hspace{-0.3mm}+\hspace{-0.3mm}5(s^2t+st^2)\hspace{-0.3mm}+\hspace{-0.3mm}7s^2t^2\hspace{-0.3mm}-\hspace{-0.3mm}2(s^3t^2+s^2t^3)\bigg)\,.
\end{align}

\subsection{Outline of paper}

The content of the remainder of this paper is organized as follows 
\\[2mm]
\noindent Section~\ref{Sec2} provides a prescription for computing the Brower-Goddard partial amplitudes by partitioning the higher-dimensional Koba-Nielsen integral according to values of conformal cross ratios and argues for the necessity of crossing symmetry.
\\[2mm]
\noindent Section~\ref{Sec3} studies the partial amplitude decomposition for the special case $d=2$. The integral over the sphere that produces the bosonic closed string amplitude is carved into three domains, and the evidence is reviewed for why each of these sub-integrals should likely be interpreted as an individual partial amplitude.
\\[2mm]
\noindent Section~\ref{Sec4} provides a recursion formula by which a $(d+2)$-dimensional partial amplitude can be immediately computed from the $d$-dimensional amplitude and states the generalization of equation \eqref{d2sumFormula} to any $d$.
\\[2mm]
\noindent 
\\[2mm]
\noindent 

\section{\label{Sec2}Partial Hyperamplitudes}

We seek to answer the question whether the $d$-dimensional uplift of Veneziano into the Brower-Goddard dual models admits a decomposition into physically sensible partial amplitudes: 
\begin{align}
\label{Adsum}
A^{(d)}(s,t) = A_p^{(d)}(s,t)+A_p^{(d)}(t,u)+A_p^{(d)}(u,s)\,.
\end{align}
For the open string, the partial amplitudes have a simple interpretation in that they are associated to different orderings of vertex operator insertions on the boundary of the disk, and on fixing the moduli space by choosing the locations of three insertions, the remaining integral naturally decomposes in three:
\begin{align}
\nonumber
&\hspace{29mm}
\begin{matrix}
\text{
\scalebox{0.8}{
\begin{tikzpicture}
\begin{scope}
\clip
(-1.1,-1.1) rectangle (0,1.1);
\draw[thick,green] (0,0) ellipse (1cm and 1.cm);
\end{scope}
\begin{scope}
\clip
(0,0) rectangle (1.1,1.1);
\draw[thick,red] (0,0) ellipse (1cm and 1.cm);
\end{scope}
\begin{scope}
\clip
(0,-1.1) rectangle (1.1,0);
\draw[thick,blue] (0,0) ellipse (1cm and 1.cm);
\end{scope}
\filldraw[white] (1,0) circle (2.8pt);
\filldraw[black] (1,0) circle (1.8pt);
\filldraw[white] (0,1) circle (2.8pt);
\filldraw[black] (0,1) circle (1.8pt);
\filldraw[white] (0,-1) circle (2.8pt);
\filldraw[black] (0,-1) circle (1.8pt);
\node at (0.05,1.4) {$\theta_1$};
\node at (0.05,-1.4) {$\theta_4$};
\node at (1.4,0) {$\theta_3$};
\end{tikzpicture}}
}
\end{matrix}
\\  \nonumber
&A^{(1)}(s,t) \hspace{-0.4mm}=\hspace{-0.4mm}
|e^{i\theta_1}\hspace{-0.4mm}-\hspace{-0.4mm}e^{i\theta_3}|\,
|e^{i\theta_1}\hspace{-0.4mm}-\hspace{-0.4mm}e^{i\theta_4}|\,
|e^{i\theta_3}\hspace{-0.4mm}-\hspace{-0.4mm}e^{i\theta_4}|\, \times
\\ \nonumber
&
\bigg(
\displaystyle\int_{\theta_1}^{\theta_3}
\hspace{-0.8mm}+\hspace{-0.8mm}\displaystyle\int_{\theta_3}^{\theta_4}
\hspace{-0.8mm}+\hspace{-0.8mm}\displaystyle\int_{\theta_4}^{\theta_1+2\pi}
\bigg)d\theta_2
\prod_{a=1}^3\prod_{b=a+1}^4
|e^{i\theta_a}\hspace{-0.4mm}-\hspace{-0.4mm}e^{i\theta_b}|^{2k_a\cdot k_b}
 \\[-2mm]
 \\[-2mm]  \nonumber
&=
\frac{\Gamma(-s-1)\Gamma(-t-1)}{\Gamma(-s-t-2)}
+
\frac{\Gamma(-t-1)\Gamma(-u-1)}{\Gamma(-t-u-2)}
\\ \nonumber
& \hspace{18mm}
+
\frac{\Gamma(-s-1)\Gamma(-u-1)}{\Gamma(-s-u-2)}\,,
\end{align}
where the integral over each of the three circular sections produces its own beta function in the last line. It is not a priori clear that such a decomposition into three partial amplitudes applies for the Brower-Goddard models for higher $d$. As a mathematical starting point for investigating this question, not using crossing symmetry as an input, one may ask, what is the largest number of distinct terms into which the Brower-Goddard amplitude $A^{(d)}(s,t)$ can be split such that each term is given by the same function $A_\mathcal{B}^{(d)}(s,t)$ but evaluated at different arguments? The answer to this question turns out to be six:
\begin{align}
A^{(d)}(s,t) = \,&
A_\mathcal{B}^{(d)}(s,t)+
A_\mathcal{B}^{(d)}(s,u)+
A_\mathcal{B}^{(d)}(t,s)+
\nonumber\\[-2.5mm]
\label{ABsplit}
\\[-2.5mm] \nonumber &
A_\mathcal{B}^{(d)}(t,u)+
A_\mathcal{B}^{(d)}(u,s)+
A_\mathcal{B}^{(d)}(u,t)\,.
\end{align}
This split into six amplitude pieces essentially amounts to carving up the integration domain of \eqref{intgeneral} into six sub-domains related by permutations of the external momenta. One way to obtain the decomposition \eqref{ABsplit} is to use the conformal symmetry as guiding principle. Given four vectors $\vec{x}_1$ to $\vec{x}_4$, one can write down a total of six conformally invariant cross ratios,
\begin{align}
&\frac{|\vec{x}_{1,2}|\,|\vec{x}_{3,4}|}{|\vec{x}_{1,3}|\,|\vec{x}_{2,4}|}\,,
\hspace{4mm}
\frac{|\vec{x}_{1,3}|\,|\vec{x}_{2,4}|}{|\vec{x}_{1,2}|\,|\vec{x}_{3,4}|}\,,
\hspace{4mm}
\frac{|\vec{x}_{1,2}|\,|\vec{x}_{3,4}|}{|\vec{x}_{1,4}|\,|\vec{x}_{2,3}|}\,,
\nonumber  \\[-2.5mm]
\\[-2.5mm] \nonumber &
\frac{|\vec{x}_{1,4}|\,|\vec{x}_{2,3}|}{|\vec{x}_{1,2}|\,|\vec{x}_{3,4}|}\,,
\hspace{4mm}
\frac{|\vec{x}_{1,3}|\,|\vec{x}_{2,4}|}{|\vec{x}_{1,4}|\,|\vec{x}_{2,3}|}\,,
\hspace{4mm}
\frac{|\vec{x}_{1,4}|\,|\vec{x}_{2,3}|}{|\vec{x}_{1,3}|\,|\vec{x}_{2,4}|}\,,
\end{align}
where $\vec{x}_{i,j}\equiv \vec{x}_i-\vec{x}_j$. Of these six, only two are independent, but for generic values of the vectors, the cross ratios assume six numerically distinct values. In consequence, one can partition the integration domain for the four-point Brower-Goddard amplitude into six sub-domains according to which cross ratio assumes the largest value. To this end, one can define a subdomain
\begin{align}
\label{Bdef}&
\mathcal{B}[abcd]
\equiv \Big\{\hspace{1mm}\vec{x}_1,\vec{x}_2,\vec{x}_3,\vec{x}_4\in\mathbb{R}^d\hspace{2mm}\Big|
\\ \nonumber
& \hspace{19mm} \frac{|\vec{x}_{a,c}|\,|\vec{x}_{b,d}|}{|\vec{x}_{a,b}|\,|\vec{x}_{c,d}|} 
\text{ is the biggest cross-ratio}
\hspace{1mm}
\Big\}\,.
\end{align}
Restricting the integral \eqref{intgeneral} to such a subdomain produces a function $A_\mathcal{B}^{(d)}(s,t)$ for which \eqref{ABsplit} applies and for which there are poles only in one channel and no crossing symmetry:
\begin{align}
A_\mathcal{B}^{(d)}(s,t)= \int_{\mathcal{B}[1234]}  d\Omega_4^{(d)}\,\prod_{i=1}^{3}\prod_{j=i+1}^4|\vec{x}_j-\vec{x}_i|^{2k_i\cdot k_j} \,.
\end{align}
This integral can be evaluated analytically, at least for odd values of $d$, where the first few cases give
\begin{align}
\label{AB1}
&A^{(1)}_\mathcal{B}(s,t)
=-\frac{{}_2F_1\big[\{-s\hspace{-0.5mm}-\hspace{-0.5mm}1,-s\hspace{-0.5mm}-\hspace{-0.5mm}t\hspace{-0.5mm}-\hspace{-0.5mm}2\};\{-s\};-1 \big]}{s+1},
\\
&A^{(3)}_\mathcal{B}(s,t)
=
\frac{2\pi}{(s+4)(s+t+8)}
\times
\\ \nonumber
&
\hspace{4mm}
\bigg(
1-{}_2F_1\Big[
\big\{-s-4,-s-t-8
\big\};
\big\{-s-3
\big\};-1
\Big]
\bigg)\,,
\\
&A_\mathcal{B}^{(5)}(s,t)
=
\frac{\pi^2}{2(s+8)}
\bigg(
\frac{4(s+7)}{(s+6)(s+t+14)}
\\ \nonumber
&\hspace{9mm}
-\frac{8}{(s+4)(s+6)(s+t+12)}
-\frac{1}{s+t+16}
\bigg)
\\ \nonumber
&+4\pi^2\frac{(s+t+15)^{-1}(s+t+16)^{-1}}{(s+t+12)(s+t+14)}\bigg(
2^{s+t+15}(t-s)-
\\ \nonumber
&
\frac{(s+7)(t+7)}{(s+8)}
{}_2F_1\big[\{-s\hspace{-0.5mm}-\hspace{-0.5mm}8,-s\hspace{-0.5mm}-\hspace{-0.5mm}t\hspace{-0.5mm}-\hspace{-0.5mm}16\};\{-s\hspace{-0.5mm}-\hspace{-0.5mm}7\};-1\big]
\bigg).
\end{align}
To obtain a standard partial amplitude, one can add together two of the smaller pieces:
\begin{align}
A_p^{(d)}(s,t) = A_\mathcal{B}^{(d)}(s,t)+A_\mathcal{B}^{(d)}(t,s)\,,
\end{align}
so that \eqref{Adsum} follows immediately from \eqref{ABsplit}, and crossing symmetry $A_p^{(d)}(s,t)=A_p^{(d)}(t,s)$ is readily apparent. 
\\[2mm]
It may be noted that in the case $d=1$ a one-parameter generalization of the identity that splits the Veneziano amplitude into two copies of \eqref{AB1} with separate sums over $s$-poles and over $t$-poles was given in \cite{Saha:2024qpt}, who interpreted their formula from a string field theory point of view as representing the freedom to perform field redefinitions that introduce four-point contact terms in addition to $s$- and $t$-exchange diagrams. The generalization was proven mathematically in \cite{Rosengren:2024rpx}; see also \cite{Bhat2025:stringy}.
\\[2mm]
Last year saw the publication of \cite{Copetti:2024rqj}, which studied integrable deformations of minimal models and demonstrated that, contrary to earlier expectations, the $S$-matrices for the kinks these deformations flow to do in fact not exhibit standard crossing symmetry. While this study applied to the specific case of quantum field theories in $D=2$ spacetime dimensions, one could imagine, in the present context, dressing the terms $A_\mathcal{B}^{(d)}(s,t)$ and $A_\mathcal{B}^{(d)}(t,s)$ with differing coefficients in an attempt to generalize Chan-Paton factors and string scattering to situations without crossing symmetry. But this does not appear to be the right path to tread. For in the limit of large $s$ and fixed $t$, the amplitude pieces grow exponentially:
\begin{align}
\nonumber
&A^{(1)}_\mathcal{B}(s,t) \sim -\frac{2^{s+t+3}}{s}\,,
\hspace{8mm}
A^{(3)}_\mathcal{B}(s,t) \sim -\frac{2^{s+t+10}\pi}{s^2}\,,
\\
&A^{(5)}_\mathcal{B}(s,t) \sim -\frac{2^{s+t+17}\pi^2}{s^3}\,.
\end{align}
To cancel out this pathological high-energy behaviour, the amplitude pieces $A^{(1)}_\mathcal{B}(s,t)$ and $A^{(1)}_\mathcal{B}(t,s)$ must be added together with identical coefficients. We can think of this fact as representing a derivation of crossing symmetry in this context and justifying the three-fold decomposition of \eqref{Adsum}, with the partial amplitudes given by
\begin{align}
\label{A}
A^{(d)}_p(s,t)\hspace{-0.5mm}=\hspace{-1.5mm}\int_{\mathcal{B}[1234]\cup \mathcal{B}[2341]}
\hspace{-13mm}
\begin{matrix}d\Omega_4^{(d)}
\\[2mm]
\end{matrix}
\hspace{6mm}\prod_{i=1}^{3}\prod_{j=i+1}^4|\vec{x}_j-\vec{x}_i|^{2k_i\cdot k_j} .
\end{align}
Performing the gauge-fixing choice $\vec{x}_1=0$, $\vec{x}_2=\hat{e}_1$, and $\vec{x}_4=\infty$ and carrying out the angular integration, one arrives at the formula
\begin{align}
\label{partialInt}
&A^{(d)}_p(s,t)
=\frac{\pi^{\frac{d-1}{2}}}{\Gamma(\frac{d-1}{2})}\int_0^{1/2}dx \int_0^{x(2-x)}dv\,v^{\frac{d-3}{2}} \,\times
\\ \nonumber
& \hspace{17mm}
\big(x^2+v\big)^{-d-\frac{s}{2}}
\big((1-x)^2+v\big)^{-d-\frac{t}{2}}
+(s\leftrightarrow t)\,.
\end{align}
This formula was previously given in \cite{Bjerrum-Bohr:2024wyw}, which also provided a closed-form answer for the odd-$d$ partial amplitudes:
\begin{align}
\label{oddDformula}
&A^{(d)}_p(s,t)
=-(16\pi)^{\frac{d}{2}}
\frac{\Gamma(-s\hspace{-1mm}-\hspace{-1mm}2d+2)\Gamma(-t\hspace{-1mm}-\hspace{-1mm}2d+2)}{\sqrt{1024\pi}\,\Gamma(-s-t-2d-1)}
\\ \nonumber & 
\times \big(\frac{s+d+2}{2}\big)_{\frac{d-3}{2}}
\big(\frac{t+d+2}{2}\big)_{\frac{d-3}{2}}
\big(\frac{s+t+2d+3}{2}\big)_{\frac{d-3}{2}}
\\ \nonumber
&-\frac{(-\pi)^{\frac{d-1}{2}}}{2^{d-4}\Gamma(\frac{d-1}{2})}
\bigg(
\sum_{M=0}^{\frac{d-3}{2}}
\frac{4^M\left(\frac{3-d}{2}\right)_M}{M!(s+4+2M)(t+4+2M)}-
\\[-2mm] \nonumber
&
4
\sum_{M=0}^{\frac{d-5}{2}}
\frac{(3-d)_M}{M!\left(\frac{4-d}{2}\right)_M(s+2d-2-2M)(t+2d-2-2M)}
\\ \nonumber
&
\times\hspace{-4mm}
\sum_{n=0}^{d-4-2M}
\frac{(d-n-3-2M)^2\left(\frac{4-d+2n}{2}\right)_M(3+M-d)_n}{n!(s+4+2M+2n)(t+4+2M+2n)}
\bigg).
\end{align}
For the first three cases, the functions are given explicitly by 
\begin{align}
&A_p^{(1)}(s,t) = \frac{\Gamma(-s-1)\Gamma(-t-1)}{\Gamma(-s-t-2)},
\\[1mm]
&A_p^{(3)}(s,t) =\hspace{-0.5mm} -\frac{2\pi\Gamma(-s\hspace{-0.7mm}-\hspace{-0.7mm}4)\Gamma(-t\hspace{-0.7mm}-\hspace{-0.7mm}4)}{\Gamma(-s-t-7)}\hspace{-1mm}+\hspace{-1mm}\frac{2\pi}{(s\hspace{-0.7mm}+\hspace{-0.7mm}4)(t\hspace{-0.7mm}+\hspace{-0.7mm}4)},
\\[1mm]
&A_p^{(5)}(s,t) = 
\\ & \nonumber
-\frac{4\pi^2(s+7)(t+7)(s+t+13)\Gamma(-s-8)\Gamma(-t-8)}{\Gamma(-s-t-11)}+
\\[-2mm] \nonumber
& 
\frac{\pi^2/2}{\prod_{n=2}^4(s\hspace{-0.7mm}+\hspace{-0.7mm}2n)(t\hspace{-0.7mm}+\hspace{-0.7mm}2n)}\Big(2240+928(s+t)+80(s^2+t^2)
\\ \nonumber & \hspace{31mm}
+388st+34(s^2t+st^2)+3s^2t^2\Big).
\end{align}
It may be observed that the meromorphic first term on each right-hand side contains non-polynomial poles for $d>1$ but that these are precisely cancelled by the subsequent rational piece. 
\\[2mm]
The following two sections will in turn study the partial amplitudes for $d=2$ and even $d>2$, respectively. As for the odd-$d$ partial amplitudes \eqref{oddDformula}, it may be remarked that they bear some relation to recent studies in the context of the $S$-matrix bootstrap. For the $d=3$ partial amplitude turns out to equal a special case ($r=1$) of the Cheung-Remmen hypergeometric candidate amplitude introduced in \cite{Cheung:2023adk}, and also studied in \cite{Rigatos:2023asb,Mansfield:2024wjc}, up to an overall shift of Mandelstam invariants, amounting to a different choice of Regge intercept:
    \begin{align}
\nonumber
&
\frac{\Gamma(-s)\Gamma(-t)}{\Gamma(-s-t)}{}_3F_2\Big[
\big\{\hspace{-1mm}-\hspace{-0.5mm}s,-t,r\big\};
\big\{\hspace{-1mm}-s\hspace{-0.5mm}-\hspace{-0.5mm}t,1+r\big\};
1
\Big]\Big|_{r=1}
\hspace{-0.5mm}=
\\    \label{r1amp}
&
\frac{1}{(s+1)(t+1)}
\hspace{-0.5mm}-\hspace{-0.5mm}\frac{\Gamma(-s\hspace{-0.5mm}-\hspace{-0.5mm}1)\Gamma(-t\hspace{-0.5mm}-\hspace{-0.5mm}1)}{\Gamma(-s-t-1)}
\hspace{-0.5mm}=\hspace{-0.5mm}\frac{A_p^{(3)}(s-3,t-3)}{2\pi}
\,.
\end{align}
Meanwhile the other partial amplitudes $A_p^{(d)}(s,t)$ do not fall into the Cheung-Remmen hypergeometric family. 
\\[2mm]
Another point of intersection with recent $S$-matrix literature concerns the condition of ``level truncation", which was identified in \cite{Cheung:2024uhn,Cheung:2024obl} as a useful bootstrap principle. The condition consists in the stipulation that for a given amplitude $A(s,t)$, there exists an infinite set of values $S=\{s_1,s_2,s_3,...\}$ such that for any $s_n\in S$, the function $f_n(t) \equiv A(s_n,t)$ only has a finite set of poles. In the case of the Veneziano amplitude $A_p^{(1)}(s,t)$, this principle is satisfied with
\begin{align} \nonumber
&S=\{-2,-3,-4,...\}\,,
\hspace{5mm}
f_1(t)=-\frac{1}{t+1}\,,
\\[-2mm]
\label{VenezianoTruncation}
\\[-2mm] \nonumber
&f_2(t)=\frac{1}{(t+1)t}\,,
\hspace{13mm}
f_3(t)=-\frac{2}{(t+1)t(t-1)}\,,
...
\end{align}
This principle is in fact satisfied for all odd-$d$ partial amplitudes $A_p^{(d)}(s,t)$. These functions, however, did not appear in \cite{Cheung:2024uhn} because the paper focused on a particular pattern of level truncation where, letting $\mathcal{P}_n$ denote the set of poles of the function $f_n(t)$,
\begin{align}
\label{simpleTruncation}
P_1 \subset P_2 \subset P_3 \subset ...\,,
\end{align}
and this pattern only applies for $A_p^{(1)}(s,t)$ and $A_p^{(3)}(s,t)$. But like the Veneziano deformations identified in \cite{Cheung:2024uhn}, the partial amplitudes $A_p^{(d)}(s,t)$ are polynomially bounded at high energy, unlike the superpolynomially soft deformations presented in \cite{Haring:2023zwu}.

\section{\label{Sec3}The Virasoro-Shapiro Partial Amplitude}

The partial amplitude integral \eqref{partialInt} is somewhat difficult to evaluate analytically for even values of $d$. One might think to eschew this obstacle by turning to numerical methods, but as the integral converges only for $s,t < -d$ and must be analytically continued to other kinematic regimes, numerical integration does not provide an entirely adequate means to study these functions.
\\[2mm]
Turning to the special case of $d=2$, to get a first analytic handle on the function $A_p^{(2)}(s,t)$, we can determine it at special values of $s$ where analytic evaluation is less difficult. For just as the Veneziano amplitude in \eqref{VenezianoTruncation} and also the higher odd-$d$ partial amplitudes reduce to rational functions at special values of $s$, so too $A_p^{(2)}(s,t)$ may be observed to simplify significantly for $s\in -4-2N_0$. The first three cases are given by
\begin{align}
 \nonumber
&A_p^{(2)}(-4,t) \hspace{-0.5mm}=
\sqrt{3}\,\frac{{}_2F_1\big[\big\{1,-\frac{t}{2}\big\};\big\{\frac{1-t}{2}\big\};\frac{1}{4}\big]}{2(t+1)(t+2)}
-\frac{2\pi}{3(t+2)},
 \\[2mm]  \nonumber
&A_p^{(2)}(-6,t) \hspace{-0.5mm}=
\frac{2\sqrt{3}(t-1)}{t^2}
-\frac{4\pi(t+1)}{3t(t+2)}
 \\  \nonumber
&\hspace{8mm}
+\frac{\sqrt{3}(8+4t+t^2)}{2t^2(t+1)(t+2)}
{}_2F_1\big[\big\{1,-\frac{t}{2}\big\};\big\{\frac{1-t}{2}\big\};\frac{1}{4}\big],
\\[-1mm]
\label{Ad2values}
\\[1mm] \nonumber
&A_p^{(2)}(-8,t) \hspace{-0.5mm}=
\hspace{-0.5mm}\frac{3(-32\hspace{-0.5mm}+\hspace{-0.5mm}32t\hspace{-0.5mm}-\hspace{-0.5mm}28t^2\hspace{-0.5mm}+\hspace{-0.5mm}7t^3)}{\sqrt{3}(t-2)^2t^2}
\hspace{-0.5mm}-\hspace{-0.5mm}\frac{4\pi(-8\hspace{-0.5mm}+\hspace{-0.5mm}3t^2)}{3(t-2)t(t+2)}
 \\  \nonumber
&
+\hspace{-0.7mm}\frac{3(128\hspace{-0.5mm}+\hspace{-0.5mm}64t\hspace{-0.5mm}+\hspace{-0.5mm}20t^2\hspace{-0.5mm}+\hspace{-0.5mm}4t^3\hspace{-0.5mm}+\hspace{-0.5mm}t^4)}{2\sqrt{3}(t\hspace{-0.5mm}-\hspace{-0.5mm}2)^2t^2(t\hspace{-0.5mm}+\hspace{-0.5mm}1)(t\hspace{-0.5mm}+\hspace{-0.5mm}2)}
{}_2F_1\big[\big\{1,-\frac{t}{2}\big\};\big\{\frac{1-t}{2}\big\};\frac{1}{4}\big].
\end{align}
As is seen above, the simplification at these values is not as drastic as for odd $d$, producing expressions involving the Gaussian hypergeometric function rather than rational functions, and for this reason the condition of level truncation is not satisfied.
\\[2mm]
The general pattern of \eqref{Ad2values} persists for larger even values of $-s$, with the amplitude evaluating to a sum of two rational pieces with coefficients that contain $\sqrt{3}$ and $\pi$, and a third piece given by a rational functions times a ${}_2F_1$ hypergeometric function with the same arguments as above. However, even knowing this data for a large set of values, it is not easy to extract the general functional form of $A_p^{(2)}(s,t)$. What can however be inferred more easily is the analytic form of the residues:
\begin{align}
\label{A2residues}
& \hspace{10mm}
\underset{s=-2+n}{\text{Res}}\Big[A_p^{(2)}(s,t)\Big]
\hspace{-0.5mm}=\hspace{-0.5mm}
-\frac{(-1)^n\pi\,\Gamma(\frac{-t-2}{2})}{n!\,\Gamma(\frac{n+2}{2})}\times
\\ \nonumber
&
{}_3\widetilde{F}_2\Big[\Big\{\hspace{-1mm}-\hspace{-0.5mm}n,\,-\frac{n}{2},\,-\frac{t+n+2}{2}\Big\};
\Big\{\frac{2-n}{2},\,-\frac{t+2n+2}{2}\Big\};1\Big],
\end{align}
for $n\in\mathbb{N}_0$, and by summing over all the poles of the amplitude, we arrive at the infinite sum representation \eqref{d2sumFormula} of the Virasoro-Shapiro partial amplitude stated in the introduction. The representation \eqref{d2sumFormula} for $A_p^{(2)}(s,t)$ as an infinite sum over a regularized Gaussian hypergeometric function, which is itself a sum, suggests that perhaps $A_p^{(2)}(s,t)$ may be expressible as a type of double-sum hypergeometric function, like the Lauricella and Appell functions. But the specific sum \eqref{d2sumFormula} does not seem to conform to the form of these special functions. 
\\[2mm]
By summing over poles with residues \eqref{A2residues} symmetrically in the $s$-, $t$-, and $u$-channels, the odd-$n$ poles cancel and the familiar Virasoro-Shapiro  amplitude $A^{(2)}(s,t)$ is recovered. A detailed survey of distinct ways of re-expressing $A^{(2)}(s,t)$ as a sum over scattering channels and contact terms was presented in Appendix~D of \cite{Saha:2024qpt}, and the cancellation of spurious poles in such processes further analyzed in \cite{Bhat2025:stringy}. The summands in the present case differ in that the polynomials are somewhat more complicated and in that contact terms are absent. 
\\[2mm]
It may be noted that \eqref{d2sumFormula} provides a partial analytic continuation of \eqref{partialInt} in the sense that it enables a continuation in $s$ but not simultaneously in $t$. For while \eqref{partialInt} only converges for $s$ and $t$ both less than $-2$, \eqref{d2sumFormula} converges for any non-singular value of $s$ provided $t$ remains sufficiently negative. In fact, in the physical kinematic regime at most one Mandelstam invariant can be positive, entailing that \eqref{d2sumFormula} suffices to  numerically compute the amplitude for any physical values of the external momenta. Moreover, the given formulas can in fact be used to obtained a full analytic continuation in both $s$ and $t$ since even for unphysical values of the Mandelstam invariants, the on-shell condition requires at least one invariant to be negative, and so $A_p^{(2)}(s,t)$ can be accessed in the unphysical regime of positive $s$ and $t$ by subtracting $A_p^{(2)}(s,u)$ and $A_p^{(2)}(t,u)$ from the full amplitude $A^{(2)}(s,t)$.
\\[2mm]
As for the extra odd-$n$ poles present in $A_p^{(2)}(s,t)$ but not in $A^{(2)}(s,t)$, we will see that none of the standard $S$-matrix tests are able to rule them out as non-physical, barring the important caveat that all poles with $n< d$ are tachyonic. Let us review then the evidence, inconclusive but suggestive, for why the function $A_p^{(2)}$ might actually represent a genuine scattering amplitude carrying physical information. The evidence may be phrased in terms of a list of properties that the function possesses. The first two properties represent non-trivial hypergeometric identities at the level of equation \eqref{d2sumFormula} but are immediate consequences of the construction of the partial amplitude by carving up the integration domain according to the procedure described in Section~\ref{Sec2}:
\\[3mm]
1. \textit{The partial amplitudes sum to the full amplitude,}
\begin{align}
\label{A2decomposition}
&A_p^{(2)}(s,t)+A_p^{(2)}(t,u)+A_p^{(2)}(u,s)
\nonumber\\[-2.5mm] 
\\[-2.5mm]\nonumber
&=
\frac{\Gamma(\frac{-s-2}{2})\,\Gamma(\frac{-t-2}{2})\,\Gamma(\frac{-u-2}{2})}{\Gamma(\frac{4+s}{2})\,\Gamma(\frac{4+t}{2})\,\Gamma(\frac{4+u}{2})},
\end{align}
for $ s+t+u=-8$. Formally, this identity follows from the partition we performed of the integration domain for the full amplitude \eqref{intgeneral}. In the kinematic regime where the sum formula \eqref{d2sumFormula} converges for all three partial amplitudes, the identity can be checked numerically.
\\[3mm]
2. \textit{Double resonance:} the sum over $t$-dependent $s$-channel poles is precisely such that the amplitudes admits an identical expansion over $s$-dependent $t$-channel poles. This follows directly from crossing symmetry: $A_p^{(2)}(s,t)=A^{(2)}_p(s,t)$.
\\[3mm]
In addition to these two properties, there are other properties indicative of $A_p^{(2)}(s,t)$ truly being an amplitude, which were not input into its construction but which nonetheless emerge on carrying out the conformal integral \eqref{A}:
\\[3mm]
3. \textit{Meromorphicity:} analytically continuing the amplitude in $s$ and $t$ to the entire complex plane, the resulting function has only isolated poles.
\\[3mm]
4. \textit{Tower of states:} the poles are all simple poles situated along equally spaced semi-infinite sequences in $s$ and $t$. 
\\[3mm]
5. \textit{Polynomial residues:} as required by locality, the residue of each $s$-channel pole is a polynomial function of $t$, which follows from the fact that the regularized hypergeometric function in \eqref{A2residues} is of the special kind where the series terminates.
\\[3mm]
6. \textit{Level-spin parity:} the partial waves of the tower of residues alternate between containing only even-spin partial waves and only odd-spin partial waves. This property is shared with known string amplitudes, for which it can be thought of as a consequence of the way states are built out of raising operators with matching spin and mass contributions. For example, applying the identity $t= \frac{s+8}{2}(\cos\theta-1)$, with $\theta$ being the scattering angle in the center of mass frame, the first six residues give
\begin{align}
\nonumber
&-\hspace{-1.5mm}\underset{s=-2}{\text{Res}}[A_p^{(2)}(s,t)] \hspace{-0.5mm}= \pi\,,
\\[1mm] \nonumber
&-\hspace{-1.5mm}\underset{s=-1}{\text{Res}}[A_p^{(2)}(s,t)] \hspace{-0.5mm}= 7+2t = 7\cos\theta\,,
\\[1mm]  \nonumber
&-\hspace{-0.5mm}\underset{s=0}{\text{Res}}\hspace{0.5mm}[A_p^{(2)}(s,t)] \hspace{-0.5mm}= \frac{1}{4}\pi(t+4)^2 = 4\pi \cos^2\theta\,,
\\[1mm] \nonumber
&-\hspace{-0.5mm}\underset{s=1}{\text{Res}}\hspace{0.5mm}[A_p^{(2)}(s,t)] \hspace{-0.5mm}= \frac{1}{72}(9+2t)(157+72t+8t^2)
\\\nonumber
&\hspace{25mm}= \frac{1}{8}(162\cos^3\theta-5\cos\theta)\,,
\\[-1.5mm]  
\\[-2.5mm] \nonumber
&-\hspace{-0.5mm}\underset{s=2}{\text{Res}}\hspace{0.5mm}[A_p^{(2)}(s,t)] \hspace{-0.5mm}
= \frac{\pi}{64}(t+4)^2(t+6)^2 
\\ \nonumber
&\hspace{25mm}= \frac{\pi}{64}(625\cos^4\theta-50\cos^2\theta+1)\,,
\\[1mm] \nonumber
&-\hspace{-0.5mm}\underset{s=3}{\text{Res}}\hspace{0.5mm}[A_p^{(2)}(s,t)] \hspace{-0.5mm}=
\\ \nonumber
&
\frac{(11\hspace{-0.5mm}+\hspace{-0.5mm}2t)(128t^4\hspace{-0.5mm}+\hspace{-0.5mm}2816t^3\hspace{-0.5mm}+\hspace{-0.5mm}22672t^2\hspace{-0.5mm}+\hspace{-0.5mm}79024t\hspace{-0.5mm}+\hspace{-0.5mm}100755)}{28\,800} \hspace{-0.5mm}
\\ \nonumber &
\hspace{15mm}
=\hspace{-0.5mm} \frac{11(117128\cos^5\theta\hspace{-0.5mm}-\hspace{-0.5mm}16940\cos^3\theta\hspace{-0.5mm}+\hspace{-0.5mm}567\cos\theta)}{28\,800}.
\end{align}

7. \textit{Positivity:} Decomposing any residue into a sum of partial waves, unitarity requires the coefficients to be all non-negative. The residues of $A_p^{(2)}(s,t)$ at even values of $s$ are equal to half the residues of the full Virasoro-Shapiro amplitude, and so positivity here follows from the unitarity of the Virasoro-Shapiro amplitude. But the poles at odd values of $s$ produce new positivity constraints. Explicitly checking the first many of these poles, one finds that positivity continues to be satisfied for the values of number $D\leq 57$ of target space dimensions permitted by the even poles. \footnote{The bound $D\leq 57$ comes from the spin-8 partial wave of the $n=10$ pole.}
\\[3mm]
8. \textit{Benign high energy behaviour:} since $A_p^{(2)}(s,t)$ is meant to be but a tree-level amplitude in a perturbative expansion, in order to have a meaningful quantum theory there must exist loop corrections created by gluing together tree-contributions. But if the tree-amplitude blows up at high energies, the loop contribution will contain pathological divergences. This, however, is not the case. From the form of the $s$-dependence in \eqref{d2sumFormula}, we see that if $A_p^{(2)}(s,t)$ goes to zero at large $|t|$ for any specific value of $s$, then it does so for any non-singular value of $s$, provided the sum converges. And at the special $s$-values given in \eqref{Ad2values}, we indeed see that the amplitude decays as $1/t$ at large $|t|$. Finally, in the unphysical regime of large positive $s$ and $t$ where the sum does not converge, $A_p^{(2)}(s,t)$ remains non-pathological, as follows from the asymptotics of the full Virasoro-amplitude and of $A_p^{(2)}(s,u)$ and $A_p^{(2)}(t,u)$, for since none of these is pathological, nor is their difference.

\section{\label{Sec4}Recursion Across Dimensions}

By a straightforward sequence of variables changes, the partial amplitude integral \eqref{partialInt} can be recast as
\begin{align}
\nonumber
&A_p^{(d)}(s,t)= \frac{\pi^{\frac{d-1}{2}}}{2^{d-4}\Gamma(\frac{d-1}{2})}
\int_0^1dv_2\int_{1-v_2}^1dv_1\,
v_1^{1-2d-s}\times
\\
&\hspace{5mm}v_2^{1-2d-t}
\Big(2v_1^2+2v_2^2+2v_1^2v_2^2-v_1^4-v_2^4-1\Big)^{\frac{d-3}{2}}.
\end{align}
This form of the integral is useful because it makes manifest a recursion relation that exists between even and odd $d$ partial amplitudes respectively:

\begin{align}
&A_p^{(d+2)}(s,t) =  
\frac{\pi}{2(d-1)}
\bigg(
2A_p^{(d)}(s+2,t+4)
+
\\ \nonumber
&2A_p^{(d)}(s+4,t+2)
+2A_p^{(d)}(s+2,t+2)
-A_p^{(d)}(s,t+4)
\\ \nonumber
& \hspace{15mm}
-A_p^{(d)}(s+4,t)
-A_p^{(d)}(s+4,t+4)
\bigg)\,.
\end{align}
For odd $d$, it can be verified that the amplitudes \eqref{oddDformula} satisfy this recursive formula. For even $d$ meanwhile, the recursive formula can applied together with the formula \eqref{d2sumFormula} for the $d=2$ amplitude to obtain the partial amplitudes $A_p^{(d)}(s,t)$ for even $d$ greater than two.

By use of the recursive formula, one can arrive at a general formula valid for even and odd $d$ alike:
\begin{align}
\label{dsumFormula}
&A_p^{(d)}(s,t)=-\sum_{n=0}^\infty\frac{(-1)^n\,\pi^{d/2}\,\Gamma(-\frac{t+d}{2})}{n!\,\Gamma(\frac{n+d}{2})}\,
\times
\\ \nonumber &\hspace{2mm}
\frac{{}_3\widetilde{F}_2\Big[\Big\{-n,\,-\frac{n}{2},\,\frac{-t-n-2d+2}{2}\Big\};\Big\{\frac{2-n}{2},\,-\frac{t+2n+d}{2}\Big\};1\Big]}
{s-n+d}\,.
\end{align}
For odd $d$, this sum simplifies and can be evaluated to recover \eqref{oddDformula}, while for even $d$, the functions $A_p^{(d)}(s,t)$ are significantly more complicated. The situation is comparable to that of conformal blocks, which are simpler and more amenable to closed-form analytic evaluation in even than in odd spacetime dimensions.
\\[2mm]
The arguments listed in Section~\ref{Sec3} for supposing $A_p^{(2)}(s,t)$ to be a physical object apply equally to the higher even-$d$ functions $A_p^{(d)}(s,t)$, as they do for the odd-$d$ cases. The ranges of target space dimensions for which positivity is satisfied changes, but inspection of the first many residues indicates that the odd poles do not restrict the allow target space dimensions beyond the limitations set by the even poles and determined in \cite{Bjerrum-Bohr:2024wyw}, signifying that positivity is always satisfied when $D\leq 10$, and is satisfied for $d\leq 6$ if $D\leq 26$. 

The level-spin parity condition also continues to apply. To clarify, it is no deep fact that the even-spin partial waves at every residue of a partial amplitude exactly matches (one half times) the residues of the full amplitude and that the only new poles appearing in the partial amplitude are due to odd-spin partial waves. This follows from the simple kinematic fact that when expressed in terms of $s$ and c.o.m. scattering angle $\theta$, the $t$ and $u$ Mandelstam invariants for general values of mass $m$ are given by
\begin{align}
t = \hspace{-0.7mm}-\frac{s-4m^2}{2}(1\hspace{-0.7mm}-\hspace{-0.7mm}\cos\theta),
\hspace{1mm}
u = \hspace{-0.7mm}-\frac{s-4m^2}{2}(1\hspace{-0.7mm}+\hspace{-0.7mm}\cos\theta),
\end{align}
so that odd powers of $\cos\theta$ cancel in any function symmetric in $t$ and $u$. The non-trivial statement is that the additional $s$-residues present in the partial amplitude $A_p^{(d)}(s,t)$ but not in the full amplitude $A^{(d)}(s,t)$, situated only at $s\in -d+1+2\mathbb{N}_0$, are polynomials expressible as positively weighted sums of Gegenbauer polynomials in $\cos\theta$.
\\[2mm]
The main objections to made against the higher-$d$ partial amplitudes $A_p^{(d)}(s,t)$ is that they contain an increasing number of tachyons and that for $d>2$, they contain massless higher-spin particles in their spectra.

\section{\label{Sec5}Level-Spin Parity and Supersymmetry}

In the special case $d=1$, the tachyon problem of the Veneziano amplitude is known to find its cure in the massless superamplitude
\begin{align}
\label{As1}
A_s^{(1)}(s,t)= \frac{\Gamma(-s)\Gamma(-t)}{\Gamma(1-s-t)}\,.
\end{align}
One may ask if a similar cure exists for the higher-$d$ partial amplitudes. It would be desirable to determine a general prescription by which one might from a given higher-$d$ amplitude $A_p^{(d)}(s,t)$ obtain a corresponding massless superamplitude; see \cite{Siegel:2020gws} for initial steps in this direction. But this section will merely investigate whether functions exist satisfying the properties expected of such partial superamplitudes. If they exist, their expansions at small values of $s$ and $t$ should assume the general form exhibited by a superamplitude,
\begin{align}
A(\epsilon s,\epsilon t)
=\,&
\frac{a_{-2,0}}{st}\,\frac{1}{\epsilon^2}
\hspace{-0.3mm}-\hspace{-0.3mm}a_{0,0}
\hspace{-0.3mm}-\hspace{-0.3mm}a_{1,0}(s+t)\,\epsilon
\nonumber
\\  \label{EFTexpansion}
&-\hspace{-0.5mm}\Big(a_{2,0}(s^2\hspace{-0.3mm}+\hspace{-0.3mm}t^2)\hspace{-0.3mm}+\hspace{-0.3mm}a_{2,1}st\Big)\epsilon^2
\\ \nonumber
&-\hspace{-0.5mm}\Big(a_{3,0}(s^3\hspace{-0.3mm}+\hspace{-0.3mm}t^3)\hspace{-0.3mm}+\hspace{-0.3mm}a_{3,1}(s^2t\hspace{-0.3mm}+\hspace{-0.3mm}st^2)\Big)\epsilon^2
\hspace{-0.3mm}+\hspace{-0.3mm}\mathcal{O}(\epsilon^3)\,.
\end{align}
There are manifold ways in which one could modify by hand the partial amplitudes $A_p^{(d)}(s,t)$ to obtain functions with such expansions. One way to greatly reduce the space of candidate amplitudes is to impose spin-level parity. For a massless amplitude, this condition amounts to the stipulation that if we set $t=\frac{s}{2}(\cos\theta-1)$, then the residues of poles at $s\in 2\mathbb{N}$ are odd polynomials in $\cos\theta$, while the residues of poles at $s\in 1+2\mathbb{N}_0$ are even polynomials in $\cos\theta$.
\\[2mm]
Motivated by the form of the odd-$d$ tachyonic  partial amplitudes, the class of functions one can consider in searching for potential supersymmetric versions are functions given by a meromorphic piece built of gamma functions plus a rational piece. Meanwhile,  inspired by the even-$d$ functions $A_p^{(d)}(s,t)$ one could explore more complicated functions involving hypergeometric functions, but this paper will refrain from so doing.
\\[2mm]
In investigating possible variations of \eqref{As1} that preserve level-spin parity, a particular combination of gamma functions that suggests itself is the following, where $\eta\in \mathbb{N}$,
\begin{align}
\label{Fdef}
F_\eta(s,t) \equiv -(-1)^\eta\frac{\Gamma(-s-\eta+1)\Gamma(-t-\eta+1)}{\Gamma(\eta-s-t)}\,.
\end{align}
The residues of this function are given below, with $n\in \mathbb{N}_0$, where the second equation uses Euler's reflection formula and reexpresses $t$ in terms of the c.o.m. scattering angle,
\begin{align}
&\hspace{5mm}\underset{s = -\eta+1+n}{\text{Res}}\Big[ F_\eta(s,t) \Big] = \frac{\Gamma(t+n-2\eta+2)}{n!\,\Gamma(t+\eta)}
\\ \nonumber
=\,& \frac{\Gamma(\frac{3+n-3\eta}{2}+\frac{n+1-\eta}{2}\cos\theta)\,\Gamma(\frac{3+n-3\eta}{2}-\frac{n+1-\eta}{2}\cos\theta)}{\pi\,n!}
\times
\\
&\hspace{5mm}\sin\Big(\pi\frac{3\eta-n+1}{2}+\pi \frac{n+1-\eta}{2}\cos\theta\Big)
\,. \nonumber
\end{align}
The point to observe from the above equation is that when $3\eta-n+1$ is even, the residue is an odd function of $\cos\theta$ and when $3\eta-n+1$ is odd, the residue is an even function. In other words, the family of functions given in \eqref{Fdef} satisfy the condition of spin-level parity. (The prefactor $-(-1)^\eta$ plays no role in this and can be thought of as an overall sign convention.)
\\[2mm]
In the case $\eta=1$, $F_\eta(s,t)$ simply reduces to the superamplitude $A_s^{(1)}(s,t)$. But for $\eta>1$, the function $F_\eta(s,t)$ is certainly not a superamplitude, for it is afflicted with two pathologies: first, the first $\eta-1$ poles are tachyonic; and second, the residues of the poles with $s\leq 2\eta-2$ are non-polynomial. However, we have witnessed how the odd-$d$ partial amplitudes in \eqref{oddDformula} are built from a meromorphic piece containing physical and unphysical poles plus a rational piece that cancels the unphysical poles of the meromorphic piece, and we can search for functions that mimic that same structure. By this line of reasoning, we are lead to the following class of putative superamplitudes:
\begin{align}
\label{AsEta}
A_s^{(\eta)}(s,t)=F_\eta(s,t)+\mathcal{R}^{(\eta)}(s,t)\,,
\end{align}
where $\mathcal{R}^{(\eta)}(s,t)$ is a rational function in $s$ and $t$ that serves to cancel unphysical poles in the meromorphic term $F_\eta(s,t)$. For $A_s^{(\eta)}(s,t)$ to stand any chance of being a physical object then, we must impose on $\mathcal{R}^{(\eta)}(s,t)$ the conditions that all tachyonic poles cancel and that $A_s^{(\eta)}(s,t)$ contains no intersecting poles in $s$ and $t$ except the $\frac{1}{st}$ piece consistent with \eqref{EFTexpansion}. These stipulations do not uniquely determine $\mathcal{R}^{(\eta)}(s,t)$, but a unique answer does exist for each $\eta$ if we adjoin one more condition, namely if we impose particularly benign high energy asymptotics by demanding that $\mathcal{R}^{(\eta)}(s,t)$ tends to zero at large values of $|s|$ and $|t|$ as  
\begin{align}
\label{largea}
\mathcal{R}^{(\eta)}(a s,a t) = \mathcal{O}(\frac{1}{a^3})\,.
\end{align}
The unique solution to $\mathcal{R}^{(\eta)}(s,t)$ obeying these conditions can be straightforwardly determined on a computer for the first many values of $\eta$. The expressions quickly become rather lengthy, but the first three instances of $\mathcal{R}^{(\eta)}(s,t)$ are given by
\begin{align}
&R^{(1)}(s,t)=0\,,
\nonumber \\ \nonumber &
R^{(2)}(s,t)=\frac{1}{6\prod_{n=-2}^1(s+n)(t+n)}
\Big(
28-14s-14t
\\ \nonumber &
-4s^2-4t^2+2s^3+2t^3-9st+5s^2t+5st^2+7s^2t^2
\\  &
-2s^3t^2-2s^2t^3
\Big)
\,,
\\ \nonumber
&R^{(3)}(s,t)=\frac{1}{360\prod_{n=-4}^2(s\hspace{-0.5mm}+\hspace{-0.5mm}n)(t\hspace{-0.5mm}+\hspace{-0.5mm}n)}\Big(
\hspace{-1mm}-\hspace{-1mm}102528 \hspace{-0.5mm}+\hspace{-0.5mm} 59808 s 
\\ \nonumber &
- 9984 s^2 + 840 s^3 + 168 s^4 - 168 s^5 + 24 s^6 + 
 59808 t
 \\ \nonumber &
 + 6008 s t  - 10052 s^2 t + 380 s^3 t - 68 s^4 t + 92 s^5 t - 
 8 s^6 t 
 \\ \nonumber &
 - 9984 t^2 - 10052 s t^2 + 450 s^2 t^2 + 130 s^3 t^2 - 
 156 s^4 t^2 
 \\ \nonumber &
 + 202 s^5 t^2 
 - 30 s^6 t^2 \hspace{-0.3mm}+\hspace{-0.3mm} 840 t^3 \hspace{-0.3mm}+\hspace{-0.3mm} 380 s t^3 \hspace{-0.3mm}+\hspace{-0.3mm}
 130 s^2 t^3 \hspace{-0.3mm}+\hspace{-0.3mm} 815 s^3 t^3 
 \\ \nonumber &
 + 100 s^4 t^3 - 115 s^5 t^3 + 10 s^6 t^3 + 
 168 t^4\hspace{-0.3mm} - \hspace{-0.3mm}68 s t^4 \hspace{-0.3mm}-\hspace{-0.3mm} 156 s^2 t^4 
 \\ \nonumber &
 + 100 s^3 t^4 
- 18 s^4 t^4  - 
 32 s^5 t^4 + 6 s^6 t^4 - 168 t^5 + 92 s t^5 
 \\ \nonumber &
 + 202 s^2 t^5 - 
 115 s^3 t^5 - 32 s^4 t^5 + 23 s^5 t^5 - 2 s^6 t^5 + 24 t^6 
 \\ \nonumber &
 - 
 8 s t^6 - 30 s^2 t^6 
+ 10 s^3 t^6 + 6 s^4 t^6 - 2 s^5 t^6
\Big)\,.
\end{align}
Although $R^{(\eta)}(s,t)$ is only constrained to eliminate the tachyonic and non-polynomial residues from $A_s^{(\eta)}(s,t)$, it turns out these constraints end up eliminating the poles at $s\in \{1,...,2\eta-2\}$ altogether. Hence, the towers of states for the functions $A_s^{(\eta)}(s,t)$ are somewhat different from those of known string amplitudes in that, for increasing $\eta$, there is an increasing gap between the ground state mass-squared and the first excited level, and only from the first excited level onward are the mass-squared values separated by unit spacing. To illustrate this behaviour and the spin-level parity phenomenon, the first five residues are listed below for $\eta=2$,
\begin{align}
&\underset{s=0}{\text{Res}}[A_s^{(2)}(s,t)]=\frac{1}{6t}\,,
\nonumber \\
&\underset{s=3}{\text{Res}}[A_s^{(2)}(s,t)]=\frac{1}{24}\,,
\nonumber \\
&\underset{s=4}{\text{Res}}[A_s^{(2)}(s,t)]=\frac{t+2}{120}=\frac{\cos\theta}{60}\,,
\\ \nonumber 
&\underset{s=5}{\text{Res}}[A_s^{(2)}(s,t)]=\frac{(t+2)(t+3)}{720}
=\frac{5}{576}\cos^2\theta-\frac{1}{2880}\,,
\\ \nonumber 
&\underset{s=6}{\text{Res}}[A_s^{(2)}(s,t)]=\hspace{-0.3mm}\frac{(t+2)(t+3)(t+4)}{5040}
\hspace{-0.3mm}=\hspace{-0.3mm}\frac{3\cos^3\theta}{560}\hspace{-0.3mm}-\hspace{-0.3mm}\frac{\cos\theta}{1680}\,.
\end{align}
and for $\eta=3$, the first five residues are given by
\begin{align}
&\underset{s=0}{\text{Res}}[A_s^{(3)}(s,t)]=\frac{1}{720t}\,,
\nonumber \\
&\underset{s=5}{\text{Res}}[A_s^{(3)}(s,t)]=\frac{1}{5040}\,,
\nonumber \\
&\underset{s=6}{\text{Res}}[A_s^{(3)}(s,t)]=\frac{t+3}{40\,320}=\frac{
\cos\theta}{13\,440}\,,
\\
&\underset{s=7}{\text{Res}}[A_s^{(3)}(s,t)]=\frac{(t+3)(t+4)}{362\,880}
=\frac{7\cos^2\theta}{207\,360}-\frac{1}{1\,451\,520}\,,
\nonumber  \\
&\underset{s=8}{\text{Res}}[A_s^{(3)}(s,t)]=\hspace{-0.3mm}\frac{(t\hspace{-0.3mm}+\hspace{-0.3mm}3)(t\hspace{-0.3mm}+\hspace{-0.3mm}4)(t\hspace{-0.3mm}+\hspace{-0.3mm}5)}{3\,628\,800}
\hspace{-0.3mm}=\hspace{-0.3mm}\frac{\cos^3\theta}{56\,700}\hspace{-0.3mm}-\hspace{-0.3mm}\frac{\cos\theta}{907\,200}.
\nonumber 
\end{align}
In addition to level-spin parity, the functions $A_s^{(\eta)}(s,t)$ also fulfill the level truncation condition of \cite{Cheung:2024uhn,Cheung:2024obl} with a truncation pattern that is in fact of the simplest kind described in \eqref{simpleTruncation}.
\\[2mm]
There are other families of similar functions with the right pole structure that too satisfy spin-level parity and level truncation, for example the following function (which does not obey \eqref{largea} but has a more standard integer spectrum):
\begin{align}
\nonumber
&\frac{(s-1)(t-1)\Gamma(-s-1)\Gamma(-t-1)}{(s+t-1)(s+t)\Gamma(-s-t-1)}
\\[-2.5mm]
\\[-2.5mm] \nonumber
&+\frac{2(st+s+t-8)}{(s-2)(s+1)(t-2)(t+1)}\,.
\end{align}
But this function and others like it are debunked once we require partial wave unitarity. Meanwhile, for the functions $A_s^{(\eta)}(s,t)$, it appears on studying the first many poles that the residues do indeed, for any reasonable value of the number $D$ of spacetime dimensions, decompose into positively weighted sums of partial waves given by Gegenbauer polynomials $C_\ell^{(\frac{D-3}{2})}(\cos\theta)$. For the familiar superamplitude $A_s^{(1)}(s,t)$ positivity is satisfied precisely for $D\leq 10$. In the case $\eta = 2$, it appears positivity is satisfied precisely for $D \leq 26$, with this bound arising from the poles both at $s=5$ and at $s=6$,
\begin{align}
\nonumber
&\underset{s=5}{\text{Res}}[A_s^{(2)}(s,t)]=
\frac{26-D}{2880(D-1)}C_0^{(\frac{D-3}{2})}(\cos\theta)
\\  & \hspace{7mm}
+\frac{5}{288(D-3)(D-1)}C_2^{(\frac{D-3}{2})}(\cos\theta)\,,
\\[2mm] \nonumber
&\underset{s=6}{\text{Res}}[A_s^{(2)}(s,t)]=
\frac{26-D}{1680(D-3)(D+1)}C_1^{(\frac{D-3}{2})}(\cos\theta)
\\  & \hspace{7mm}
+\frac{9}{280(D-3)(D-1)(D+1)}C_3^{(\frac{D-3}{2})}(\cos\theta)\,,
\end{align}
while for for higher values of $\eta$ it appears positivity is satisfied for yet higher values of $D$. The techniques developed in \cite{Arkani-Hamed:2022gsa} and \cite{Mansfield:2025gca} can likely shed light on whether these positivity properties are rigorously true. 
\\[2mm]
Still, even after also demanding partial wave unitarity, the functions $A_s^{(\eta)}(s,t)$ are not uniquely singled out. Two extra isolated counterexamples of additional functions that appear to satisfy all the usual bootstrap conditions, along with spin-level parity and level truncation, are the function
\begin{align}
\nonumber
A_s^{(\alpha)}(s,t)
\equiv\,& -\frac{(st+s^2+t^2-3)\Gamma(-s-1)\Gamma(-t-1)}{\Gamma(2-s-t)}
\\ \label{Aalpha}
&+\frac{2(st-1)}{(s-1)(t-1)s t(s+1)(t+1)}\,,
\end{align}
which, judging from its first many residues, satisfy partial unitarity for spacetime dimensions $D$ less than about 26.6, and the function
\begin{align}
\nonumber &
A_s^{(\beta)}(s,t)
\equiv
-\pi^2(s+2)(t+2)(s+t-2)\,\times
\\\nonumber &
\hspace{2mm}\Big(189-22(s^2+t^2)-22st+6(s^2t+st^2)+2(s^3t+st^3)
\\ &
+3s^2t^2+s^4+t^4\Big)\frac{\Gamma(-3-s)\Gamma(-3-t)}{8\Gamma(4-s-t)}
\\\nonumber &
-
\frac{(s+2)(t+2)\pi^2/2}{\prod_{n=-3}^3(s+n)(t+n)}\Big(540-270(s+t)-36(s^2+t^2)
\\\nonumber &
-213st+18(s^3+t^3)+129(s^2t+st^2)+12(s^3t+st^3)
\\\nonumber &
+10s^2t^2-6(s^4t+st^4)-16(s^3t^2+s^2t^3)+s^4t^2+s^2t^4
\\\nonumber &
-3s^3t^3+s^4t^3+s^3t^4\Big),
\end{align}
which, is perhaps of yet more dubious physical relevance, since it appears to satisfy partial wave unitarity only for spacetime dimensions $D$ less than about 6.06. The two examples $A_s^{(\alpha)}(s,t)$ and $A_s^{(\beta)}(s,t)$ are part of an infinite family, the other members of which appear to be ruled out by partial wave unitarity.
\\[2mm]
Very likely more instances exist of functions that satisfy the bootstrap constraints that lead to the functions $A_s^{(\eta)}(s,t)$, and more exhaustive searches, perhaps along the line of \cite{Eckner:2024ggx}, could offer a more complete picture. In particular, this section only considered simple functions built of rational pieces and gamma functions. What sets apart the functions $A_s^{(\eta)}(s,t)$ is that, like the tachyonic and super-Veneziano amplitudes and the odd-$d$ partial amplitudes, their residues factorize entirely into linear factors.

\subsection{EFT bounds}

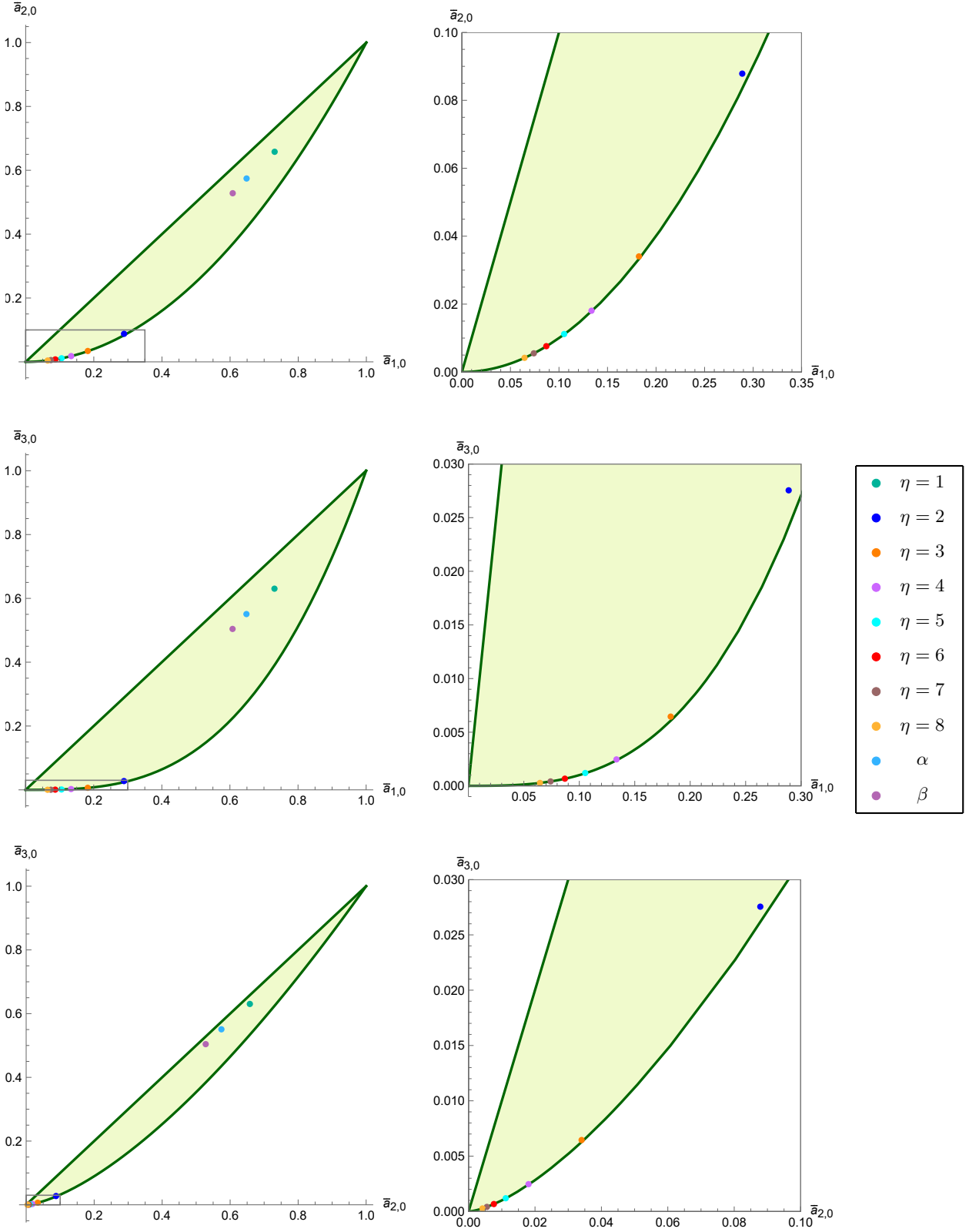
\begin{figure*}
\centering
$\hspace{-12mm}\begin{matrix}
\text{
\includegraphics[width=1.6\columnwidth]{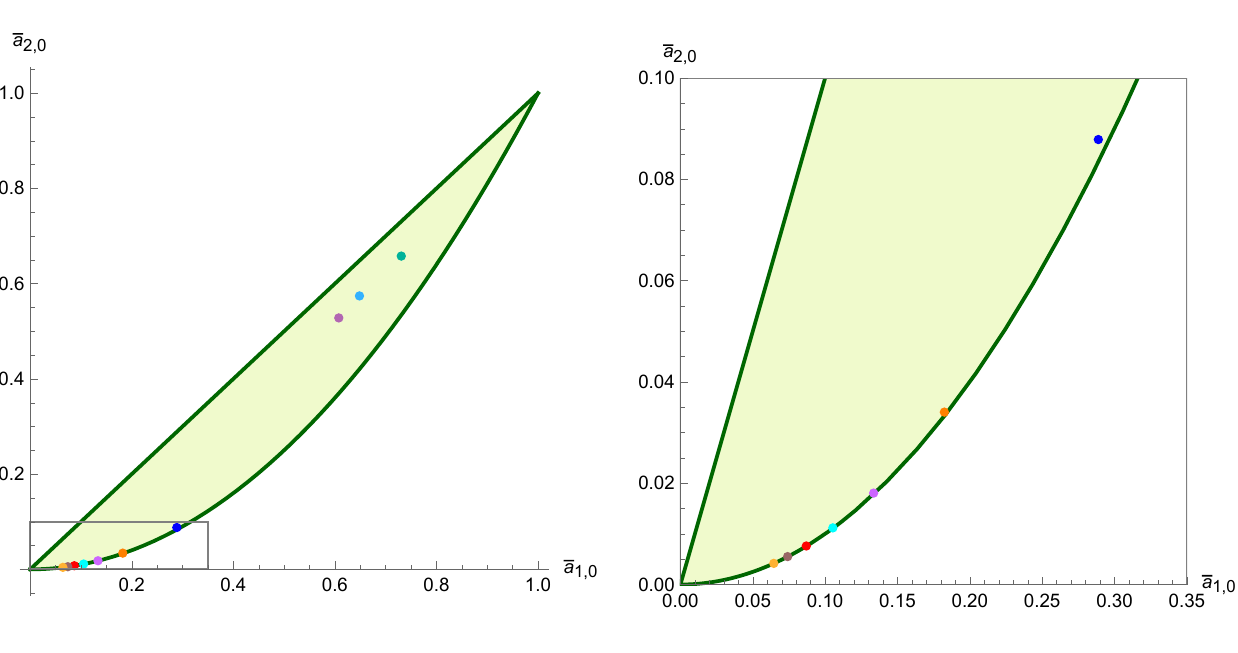}}
\\[-4mm]
\text{
\includegraphics[width=1.6\columnwidth]{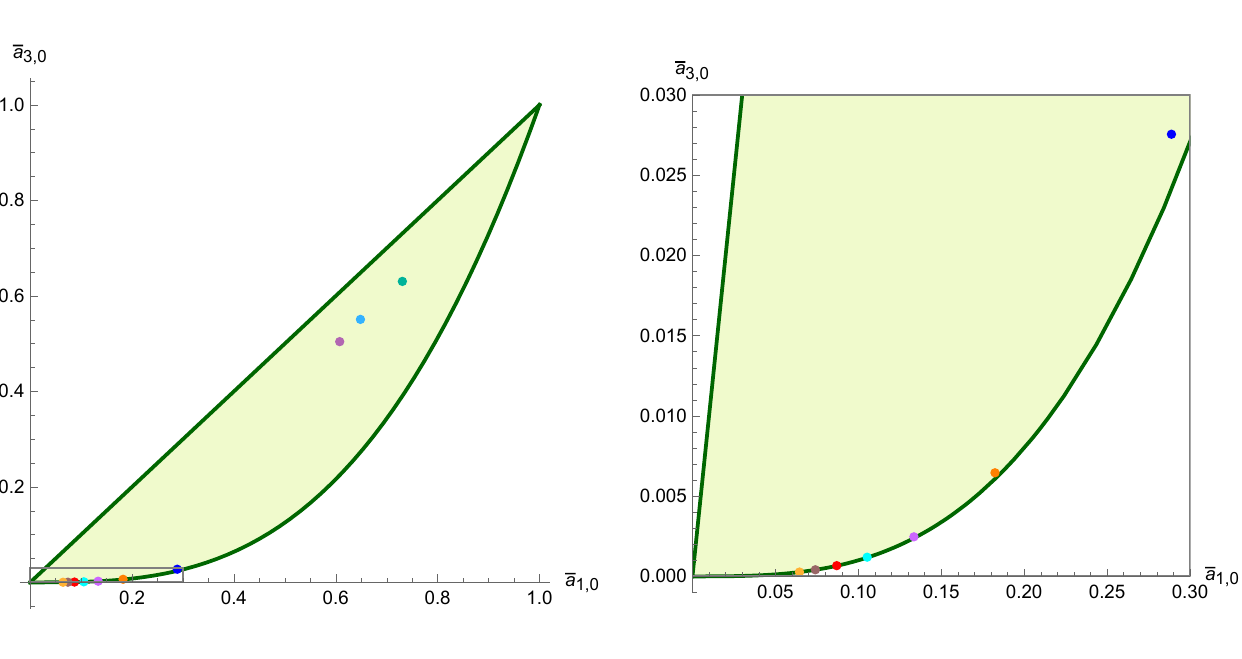}}
\\[-66mm]
\hspace{170mm}
\text{
\begin{tikzpicture}
\draw [thick](-0.35,4.5) -- (1.5,4.5);
\draw [thick](-0.35,-1.5) -- (1.5,-1.5);
\draw [thick](-0.35,-1.5) -- (-0.35,4.5);
\draw [thick](1.5,-1.5) -- (1.5,4.5);
\filldraw[colEta1] (0,4.2) circle (2pt);
\node at (0.8,4.2)  {$\eta=1$};
\filldraw[colEta2] (0,3.6) circle (2pt);
\node at (0.8,3.6)  {$\eta=2$};
\filldraw[colEta3] (0,3) circle (2pt);
\node at (0.8,3)  {$\eta=3$};
\filldraw[colEta4] (0,2.4) circle (2pt);
\node at (0.8,2.4)  {$\eta=4$};
\filldraw[colEta5] (0,1.8) circle (2pt);
\node at (0.8,1.8)  {$\eta=5$};
\filldraw[colEta6] (0,1.2) circle (2pt);
\node at (0.8,1.2)  {$\eta=6$};
\filldraw[colEta7] (0,0.6) circle (2pt);
\node at (0.8,0.6)  {$\eta=7$};
\filldraw[colEta8] (0,0) circle (2pt);
\node at (0.8,0)  {$\eta=8$};
\filldraw[colAlpha] (0,-0.6) circle (2pt);
\node at (0.8,-0.6)  {$\alpha$};
\filldraw[colBeta] (0,-1.2) circle (2pt);
\node at (0.8,-1.2)  {$\beta$};
\end{tikzpicture}
}
\\[1mm]
\text{
\includegraphics[width=1.6\columnwidth]{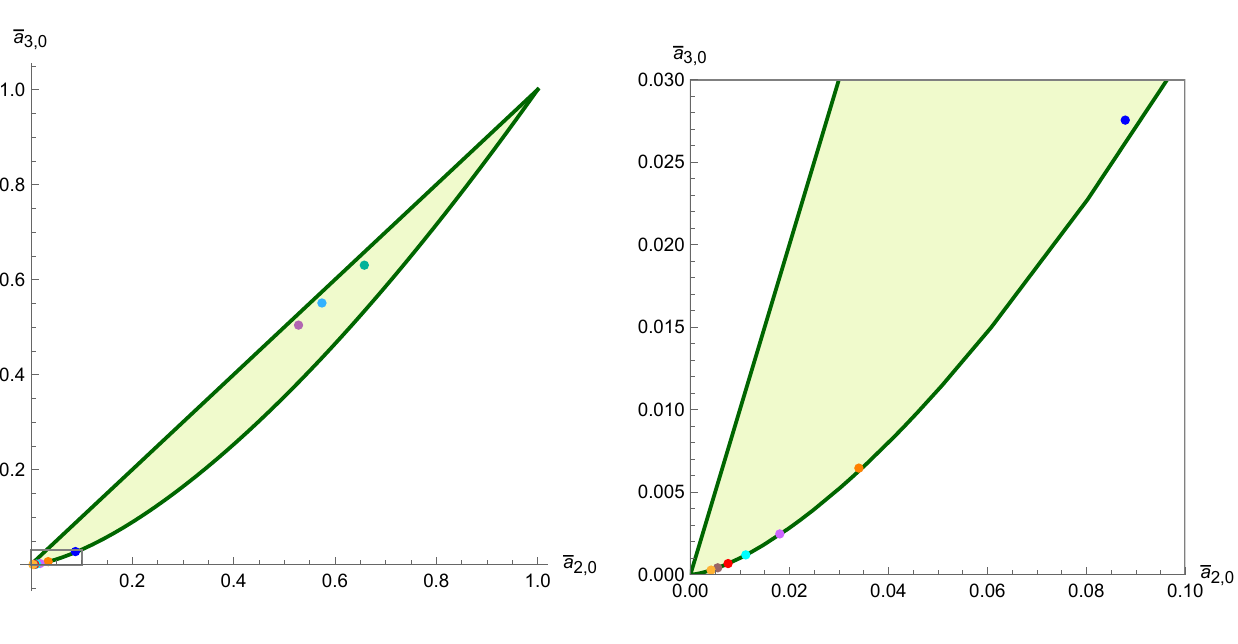}}
\end{matrix}
$
\caption{\label{fig:Islands1} 
Islands of allowed values for normalized Wilson coefficients according to the analytic bounds \eqref{analyticBounds} along with the locations of the first eight $A_s^{(\eta)}(s,t)$ functions and of $A_s^{(\alpha)}(s,t)$ and $A_s^{(\beta)}(s,t)$. The $\eta = 1$ point marks the Veneziano amplitude. The right-hand plots are zoom-ins of the left-hand ones.
} 
\end{figure*}

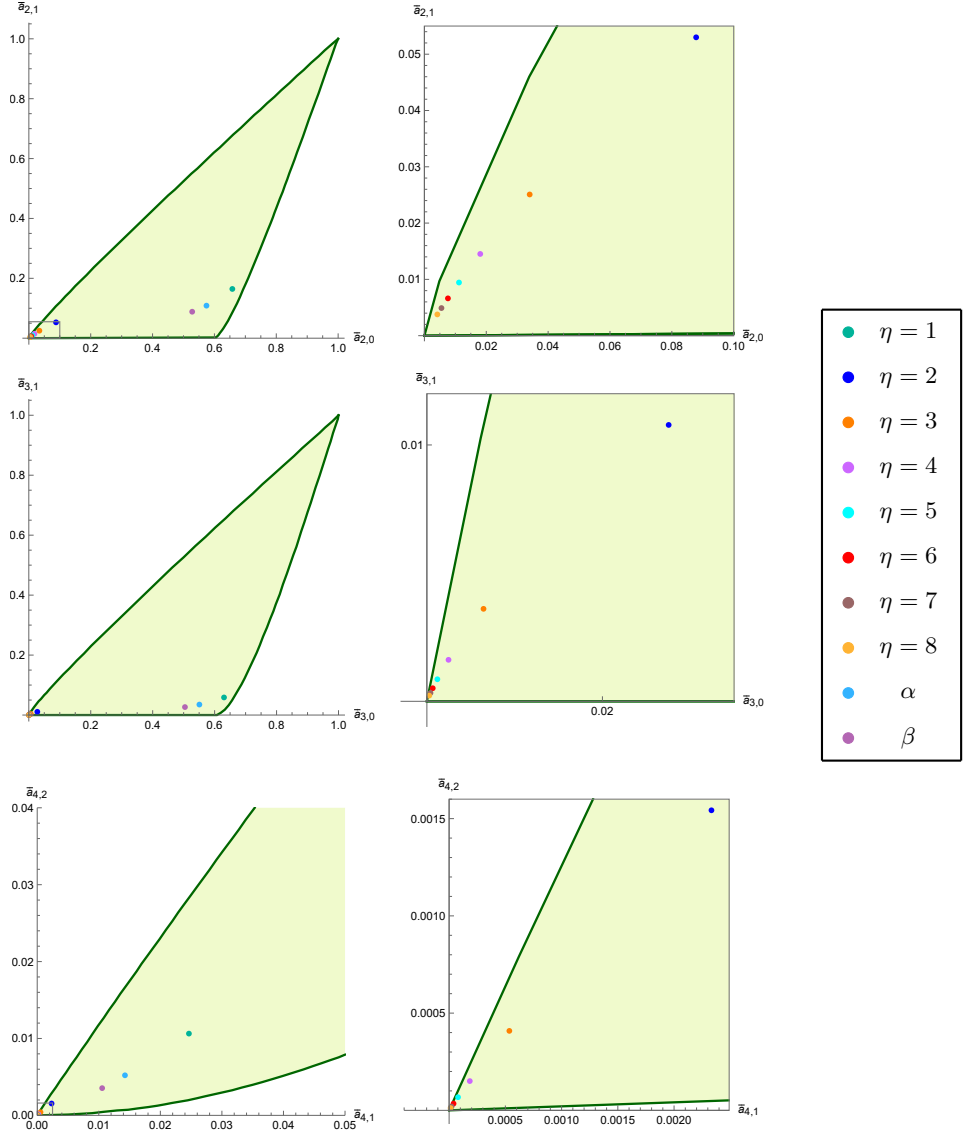
\begin{figure*}
\centering
$\hspace{-10mm}\begin{matrix}
\text{
\includegraphics[width=1.12\columnwidth]{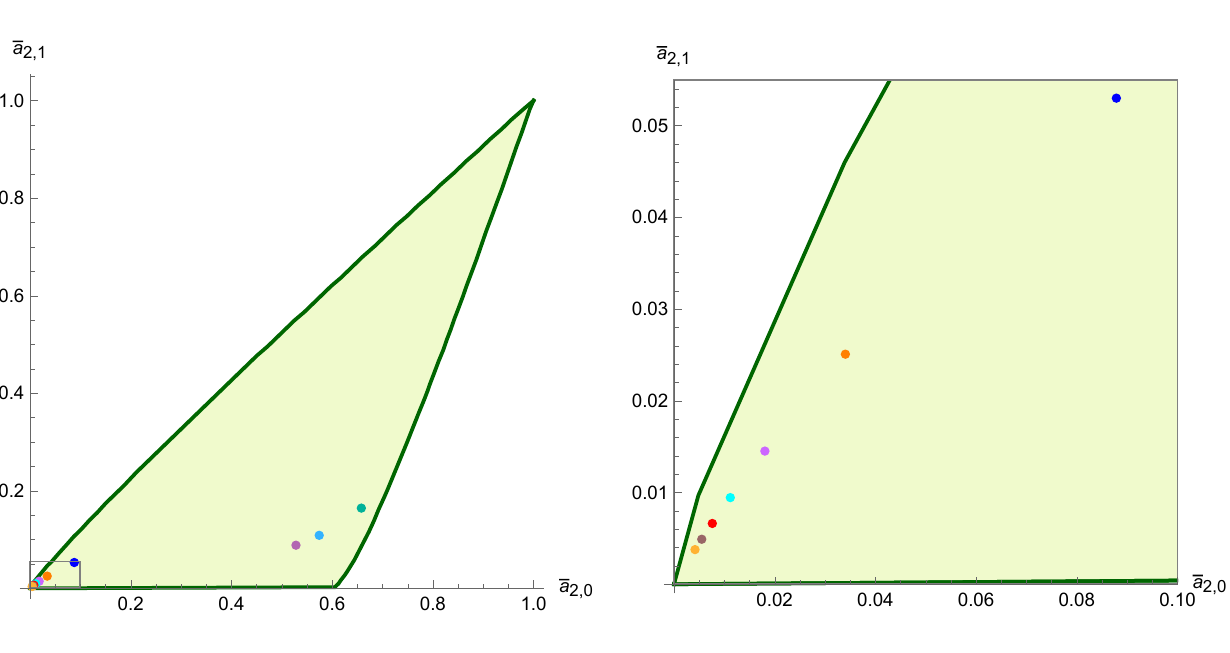}}
\\[-4mm]
\text{
\includegraphics[width=1.12\columnwidth]{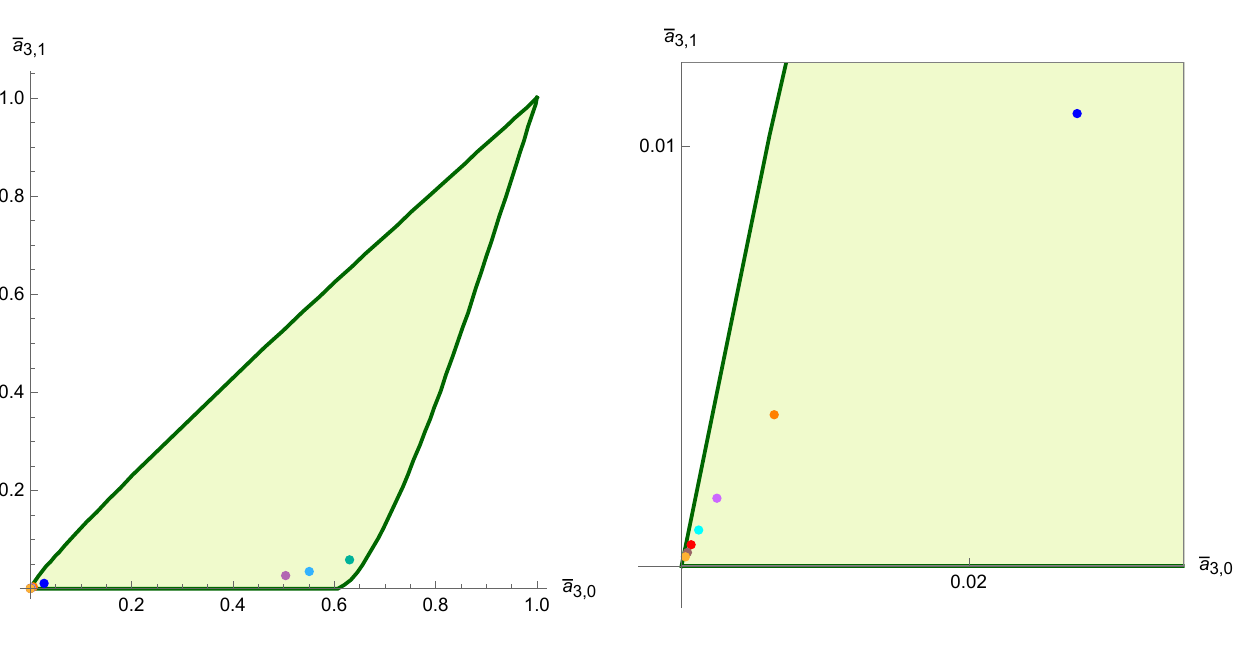}}
\\[-60mm]
\hspace{135mm}
\text{
\begin{tikzpicture}
\draw [thick](-0.35,4.5) -- (1.5,4.5);
\draw [thick](-0.35,-1.5) -- (1.5,-1.5);
\draw [thick](-0.35,-1.5) -- (-0.35,4.5);
\draw [thick](1.5,-1.5) -- (1.5,4.5);
\filldraw[colEta1] (0,4.2) circle (2pt);
\node at (0.8,4.2)  {$\eta=1$};
\filldraw[colEta2] (0,3.6) circle (2pt);
\node at (0.8,3.6)  {$\eta=2$};
\filldraw[colEta3] (0,3) circle (2pt);
\node at (0.8,3)  {$\eta=3$};
\filldraw[colEta4] (0,2.4) circle (2pt);
\node at (0.8,2.4)  {$\eta=4$};
\filldraw[colEta5] (0,1.8) circle (2pt);
\node at (0.8,1.8)  {$\eta=5$};
\filldraw[colEta6] (0,1.2) circle (2pt);
\node at (0.8,1.2)  {$\eta=6$};
\filldraw[colEta7] (0,0.6) circle (2pt);
\node at (0.8,0.6)  {$\eta=7$};
\filldraw[colEta8] (0,0) circle (2pt);
\node at (0.8,0)  {$\eta=8$};
\filldraw[colAlpha] (0,-0.6) circle (2pt);
\node at (0.8,-0.6)  {$\alpha$};
\filldraw[colBeta] (0,-1.2) circle (2pt);
\node at (0.8,-1.2)  {$\beta$};
\end{tikzpicture}
}
\\[-1mm]
\text{
\includegraphics[width=1.12\columnwidth]{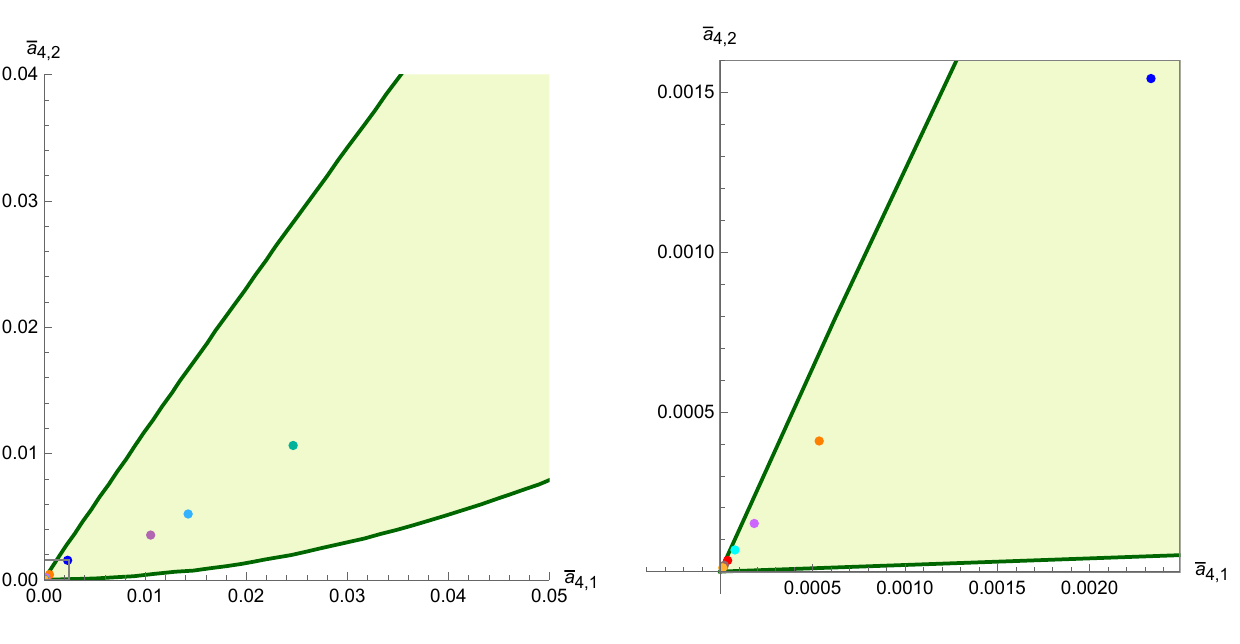}}
\\[-2mm]
\end{matrix}
$
\caption{\label{fig:Islands2} 
Islands of allowed values for normalized Wilson coefficients along with the locations of the first eight $A_s^{(\eta)}(s,t)$ functions and of $A_s^{(\alpha)}(s,t)$ and $A_s^{(\beta)}(s,t)$. The right-hand plots are zoom-ins of the left-hand ones. The islands are extracted graphically from the numerically computed regions of \cite{Berman:2023jys}.
} 
\end{figure*}

The bootstrap conditions we have considered so far still do not exhaust the known constraints that must be respected by any physical superamplitude. An important additional constraint is that the Taylor series coefficient $a_{i,j}$ in \eqref{EFTexpansion}, sometimes referred to as Wilson coefficients, are subject to bounds consequent to unitarity and locality in $4D$, which lead to finite regions of permitted values \cite{Caron-Huot:2020cmc,Arkani-Hamed:2020blm}. These bounds may conveniently be expressed in terms of normalized Wilson coefficients
\begin{align}
\overline{a}_{k,q}
\equiv \frac{a_{k,q}}{a_{0,0}}\,.
\end{align}
For some coefficients, rigorous analytic bounds are known. These include the following \cite{Arkani-Hamed:2020blm,Berman:2023jys}
\begin{align}
\nonumber &
\overline{a}_{1,0}^2
\leq 
\overline{a}_{2,0}
\leq 
\overline{a}_{1,0}\,,
\hspace{5mm}
\overline{a}_{2,0}^{3/2}
\leq 
\overline{a}_{3,0}
\leq 
\overline{a}_{2,0}\,,
\\ \label{analyticBounds}
& \hspace{18mm}
\overline{a}_{1,0}^3
\leq 
\overline{a}_{3,0}
\leq 
\overline{a}_{1,0}\,.
\end{align}
The regions in EFT-space allowed by these bounds are plotted in Figure~\ref{fig:Islands1}, where the locations of the functions $A^{(\eta)}_s(s,t)$ for $\eta\leq 8$ and of $A^{(\alpha)}_s(s,t)$ and $A^{(\beta)}_s(s,t)$ are also marked. Though it may be hard to tell from the figure since some points sit very near the boundary of the allowed regions, all points are situated inside the permitted space.
\\[2mm]
For other Wilson coefficients, bounds are not known analytically but can be computed numerically. Figure~\ref{fig:Islands2} plots the allowed regions for Wilson coefficient pairs $(\overline{a}_{2,0},\overline{a}_{2,1})$, $(\overline{a}_{3,0},\overline{a}_{3,1})$, and $(\overline{a}_{4,1},\overline{a}_{4,2})$ as extracted from the work of \cite{Berman:2023jys}, and again the Wilson coefficients of the functions $A^{(\alpha)}_s(s,t)$ and $A^{(\beta)}_s(s,t)$ and $A^{(\eta)}_s(s,t)$ for $\eta\leq 8$ are all situated inside the allowed islands. The numerical values of the Wilson coefficients for these functions are are listed in Table~\ref{tab:WilsonCoefficients}.
\\[2mm]
Within the past few years, various papers have explored bootstrap assumptions that allow one to home in on the Veneziano superamplitude in EFT-space. It may be worth remarking how the functions $A^{(\eta)}_s(s,t)$ evade the assumptions of some of these works: 
\begin{itemize}
\item The papers \cite{Chiang:2023quf,Berman:2023jys} demonstrated how the Veneziano superamplitude is virtually the only possible amplitude compatible with standard $S$-matrix bootstrap constraints and string monodromy. But the higher-$d$ amplitudes $A^{(d)}_p(s,t)$ do not model strings but rather higher-dimensional extended objects and do not satisfy string monodromy in the standard form, but rather the modified relations given in \cite{Bjerrum-Bohr:2024wyw}. And the same must be expected to apply for any supersymmetric avatars of $A^{(d)}_p(s,t)$.
\item A string-inspired input assumption of \cite{Berman:2024wyt} is that the gap between the first and second excited values of $m_n^2$ is no less than that between the ground state value $m_0^2=0$ and first excited value $m_1^2$, which as we have seen does not apply to the functions $A^{(\eta)}_s(s,t)$ with $\eta>1$ since $m_0^2=0$ and $m_1^2 = 2\eta-1$, while $m_2^2 = 2\eta$.
\end{itemize}

\begin{table*}
\caption{\label{tab:WilsonCoefficients}Numerical values of Wilson coefficients}
\begin{ruledtabular}
\begin{tabular}{ccccccccc}
\hline
$\eta$
&$a_{0,0}$ 
& $a_{1,0}$ 
&$a_{2,0}$ 
& $a_{2,1}$
&$a_{3,0}$ 
& $a_{3,1}$
&$a_{4,1}$ 
& $a_{4,2}$
\\
\hline
1
&{1.6449}
& 1.2021
&{1.0823}
& 0.27058
&{1.0369}
& 0.096551
&{0.040537}
& 0.017490
\\
\hline
2
&{0.021733}
& 0.0062764
&{0.0019094}
& 0.0011518
&{0.00059868}
& 0.00023430
&{0.000050690}
& 0.000033531
\\
\hline
3
&{0.000061203}
& $0.000011164$
&{$2.0833\cdot 10^{-6}$}
& $1.5351\cdot 10^{-6}$
&{$3.9501\cdot 10^{-7}$}
& $2.2078\cdot 10^{-7}$
&{$3.2779\cdot 10^{-8}$}
& $2.5012\cdot 10^{-8}$
\\
\hline
4
&{$6.0292\cdot 10^{-8}$}
& $8.0501\cdot 10^{-9}$
&{$1.0885\cdot 10^{-9}$}
& $8.7529\cdot 10^{-10}$
&{$1.4862\cdot 10^{-10}$}
& $9.7579\cdot 10^{-11}$
&{$1.1090\cdot 10^{-11}$}
& $9.0784\cdot 10^{-12}$
\\
\hline
5
&{$2.7214\cdot 10^{-11}$}
& $2.8666\cdot 10^{-12}$
&{$3.0440\cdot 10^{-13}$}
& $2.5698\cdot 10^{-13}$
&{$3.2534\cdot 10^{-14}$}
& $2.3406\cdot 10^{-14}$
&{$2.1596\cdot 10^{-15}$}
& $1.8428\cdot 10^{-15}$
\\
\hline
6
&{$6.6083\cdot 10^{-15}$}
& $5.7488\cdot 10^{-16}$
&{$5.0287\cdot 10^{-17}$}
& $4.3790\cdot 10^{-17}$
&{$4.4193\cdot 10^{-18}$}
& $3.3725\cdot 10^{-18}$
&{$2.6213\cdot 10^{-19}$}
& $2.2990\cdot 10^{-19}$
\\
\hline
7
&{$9.5986\cdot 10^{-19}$}
& $7.1119\cdot 10^{-20}$
&{$5.2908\cdot 10^{-21}$}
& $4.7071\cdot 10^{-21}$
&{$3.9497\cdot 10^{-22}$}
& $3.1405\cdot 10^{-22}$
&{$2.1099\cdot 10^{-23}$}
& $1.8866\cdot 10^{-23}$
\\
\hline
8
&{$8.9897\cdot 10^{-23}$}
& $5.8008\cdot 10^{-24}$
&{$3.7546\cdot 10^{-25}$}
& $3.3933\cdot 10^{-25}$
&{$2.4368\cdot 10^{-26}$}
& $1.9971\cdot 10^{-26}$
&{$1.1818\cdot 10^{-27}$}
& $1.0720\cdot 10^{-27}$
\\
\hline
$\alpha$
&{$0.93480$}
& $0.60617$
&{$0.53684$}
& $0.10161$
&{$0.51490$}
& $0.032683$
&{$0.013326$}
& $0.0048699$
\\
\hline
$\beta$
&{$0.031910$}
& $0.019392$
&{$0.016852$}
& $0.0028222$
&{$0.016089$}
& $0.00084922$
&{$0.00033663$}
& $0.00011299$
\\
\hline
\end{tabular}
\end{ruledtabular}
\end{table*}

\section{\label{Sec6}Concluding Remarks}

We have seen in this paper that the Virasoro-Shapiro closed string amplitude $A^{(2)}(s,t)$ admits a partial amplitude, in the sense that there exists a function, given in \eqref{d2sumFormula}, exhibiting crossing symmetry but not full permutation symmetry in the external momenta, which sums up to $A^{(2)}(s,t)$, and which satisfies the standard $S$-matrix requirements of polynomial residues, partial wave positivity, and polynomially bounded high energy asymptotics.

The existence of a partial amplitude of this kind is a necessary requirement for the distinguishability of closed bosonic strings in the ground state. If it is possible to endow such strings with labels akin to those provided by Chan-Paton factors in the case of open strings, whatever the origin of these labels---conceiving for example of a notion of linking involved, or imaging a closed string located on one of multiple coincident membranes---then scattering processes without full symmetry in $s$, $t$, and $u$ are possible, and functions for the associated amplitudes must exist. We now have candidates for such functions at tree level. But whether the possibility is realized, and if so which physical interpretation is the correct one, are questions which it does not seem possible to glean from amplitude formulas alone. And the objection may be raised here that the graviton sits in the closed string spectrum, and it seems unsound to introduce additional degrees of freedom if they lead to multiple gravitons \cite{Boulanger:2000bp,Boulanger:2000rq}. 

If the partial amplitudes do carry physical information beyond what is contained in the full amplitude, then it should be possible to consistently enlarge the closed bosonic string spectrum to include odd-spin states; the puzzle here is that it is unclear if the quantization of the closed string, built from integer-spin raising operators subject to a level-matching condition, can meaningfully be modified to permit such states.

We have seen too that the partial amplitude decomposition extends beyond open and closed strings to the full Brower-Goddard family of dual models, with the partial amplitudes, computed from $d$-dimensional conformal integrals, given in \eqref{dsumFormula}. The most natural way to think of these models is as describing extended objects beyond strings, but their precise interpretation remains uncertain. The only hints available at present hail from doubtful guesses as to how the known string quantizations may be uplifted. The Koba-Nielsen integral for the Veneziano amplitude is an integral not over the worldsheet of the open string but over the worldline of the string endpoints, and similarly a closed string can be thought of as the boundary of a two-dimensional topological bulk, under a perspective that straightforwardly can be applied in higher dimensions. And as the closed $d=2$ spectrum is described by two sets of coupled oscillators, it is tempting to think that the higher-$d$ cases are described by $d$ sets of oscillators which are coupled through constraints that ascend from the aforementioned level-matching condition.

In any event it is clear that if the Brower-Goddard models describe extended objects, then they describe extended objects of a particularly simple kind. Generically, the quantization of branes is expected to give rise to a continuum of states. It may be relevant to mention at this point that the type of functions studied in this paper admit $q$-deformations that posses branch cuts, i.e. continua of poles, and that such deformed functions have been identified in the $S$-matrix bootstrap as being of potential physical interest \cite{Cheung:2023adk}. While the $q$-deformation in its simplest form produces unitarity violations when applied to the open string amplitude \cite{Jepsen:2023sia}, it may have a role to play at higher $d$.

Another major simplification of the Brower-Goddard models is that the world-(hyper)volume is Euclidean. As this paper has been restricted in scope to tree-level scattering, the distinction between Euclidean and Lorentzian signature was not significant, but it plays a crucial role at loop-level, and the $i\epsilon$ prescription is already for string amplitudes a very subtle issue \cite{Witten:2013pra,Eberhardt:2023xck,Eberhardt:2024twy,Manschot:2024prc}. But even the more basic question of what the expansion parameter is currently remains unanswered as it seems a genus expansion induced by a dilaton field is not available in the present case.

Besides the signature of the worldvolume metric, one may inquire more broadly about what kinds of shapes the worldvolume is allowed to assume. The original integration domain for the Brower-Goddard model as given in \eqref{intgeneral} is $\mathbb{R}^d$, but owing to the conformal symmetry of the integral, the model can equivalently be integrated over $S^d$, which was the starting point adopted in \cite{Natsuume:1993ix}. But if the myriad possible configurations of four extended scattering objects that traverse spacetime to be summed over in a path integral can be encapsulated by such a simple worldvolume geometry, it would require an inordinate amount of symmetry, whereas the conformal group in dimensions above two is finite-dimensional. Furthermore the infinite symmetry of the extended two-dimensional conformal group plays a key role in establishing the no-ghost theorem of string theory. Without infinite symmetry then, it will be hard to make sense of worldvolume integrals as models for scattering. But on allowing deformations of the conformal structure, the higher dimensional conformal group can in fact be extended so as to incorporate an infinite-dimensional algebra \cite{Kapranov:2021nfy}, which could well play a crucial role in this story, so that there may yet be hope that the no-ghost theorem allows for a generalization.

Another open question is how to define partial amplitudes for the scattering of more than four objects. In particular, how does the manner of partitioning the worldvolume into sub-domains prescribed in \eqref{Bdef} extend to higher-point scattering? Answering this question would pave the way to checking factorization at higher-point. As work over the past years \cite{Arkani-Hamed:2023jwn,Cheung:2025krg} have made manifest just how powerful the constraining force of such higher-point checks are, subjecting the Brower-Goddard models to these checks would provide a highly reliable way of ascertaining if the models are really consistent. Over the past few years, the work of \cite{Arkani-Hamed:2023lbd} and subsequent papers have uncovered a novel way of thinking about scattering in terms of counting problems associated to curves on surfaces, and recently \cite{Cao:2025lzv} demonstrated how the superstring amplitude may be construed in this manner. Uplifting this perspective to hypersurfaces could perhaps provide a instructive tool for tackling the problem of higher-point scattering.

It may seem a moot point to raise a host of intricate questions for a class of models which are all tachyonic. But as we have witnessed in Section~\ref{Sec5} how supplementing the standard $S$-matrix bootstrap constraints with assumptions hailing from the behaviour of the higher-$d$ partial amplitudes produces discrete solutions that behave as sensible massless superamplitudes, it seems conceivable that there may well be a supersymmetric, tachyon-free side to the story. If one adopts the most pessimistic stance and supposes none of the partial amplitudes described here beyond the standard superamplitude $A_s^{(1)}(s,t)$ to be of any physical relevance, the morale of the present paper would be to underscore the insufficiency of existing amplitude bootstrap constraints. For these constraints will necessarily home in on and single out as functions of interest functions which include the kind of higher-$d$ and $\eta-$functions studied in this paper. Indeed we have already observed via \eqref{r1amp} how a bottom-up amplitude survey secretly stumbled upon the $d=3$ higher-dimensional partial amplitude.

It is possible to bootstrap the Veneziano amplitude if the standard $S$-matrix bootstrap conditions are supplemented with additional assumptions motivated by specific theories, such as the string monodromy relations \cite{Chiang:2023quf,Berman:2023jys}, and by this method no-go theorems are established within wide classes of theories. But the bootstrap approach, if it is not stripped of all assumptions not founded in the most elementary requirements of a sensible theory, loses that theory-independence which otherwise enables it to attain the crowning achievement of truly universal no-go theorems.

The opposite optimistic stance would be to believe that supersymmetric Brower-Goddard models really can serve as meaningful models for branes. But even then additional restrictions are called for to rule out more function space, as it takes perhaps an excessive stretch of the imagination to conceive of not just multiple but infinitely many distinct physically sensible UV completions of Yang-Mills theory. But one may hope for and can investigate the plausibility of an eventuality where a special subset of the kinds of models studied here touches on the very specific set of branes known to exist in string and M-theory. Certainly string theory requires branes and certainly these branes can scatter. We are still in the dark about many of the most elementary properties of branes, open and closed, and we should welcome any method that may shine new light on these objects. The $S$-matrix bootstrap may provide just such a portal.

\begin{acknowledgments}
I am grateful to Emil Bjerrum-Bohr for many enlightening discussions, to Justin Berman, Mykola Dedushenko, Sangmin Lee, Sungjay Lee, Piljin Yi, and Wayne Zhao for sharing important insights, and to Justin Berman, Henriette Elvang, and Aidan Herderschee for permitting me to re-use their EFT-hedron plots. My work is supported by the Korea Institute for Advanced Study (KIAS) Grant PG095901.
\end{acknowledgments}

\bibliography{literature}

\providecommand{\noopsort}[1]{}\providecommand{\singleletter}[1]{#1}%
\begin{thebibliography}{48}%
\makeatletter
\providecommand \@ifxundefined [1]{%
 \@ifx{#1\undefined}
}%
\providecommand \@ifnum [1]{%
 \ifnum #1\expandafter \@firstoftwo
 \else \expandafter \@secondoftwo
 \fi
}%
\providecommand \@ifx [1]{%
 \ifx #1\expandafter \@firstoftwo
 \else \expandafter \@secondoftwo
 \fi
}%
\providecommand \natexlab [1]{#1}%
\providecommand \enquote  [1]{``#1''}%
\providecommand \bibnamefont  [1]{#1}%
\providecommand \bibfnamefont [1]{#1}%
\providecommand \citenamefont [1]{#1}%
\providecommand \href@noop [0]{\@secondoftwo}%
\providecommand \href [0]{\begingroup \@sanitize@url \@href}%
\providecommand \@href[1]{\@@startlink{#1}\@@href}%
\providecommand \@@href[1]{\endgroup#1\@@endlink}%
\providecommand \@sanitize@url [0]{\catcode `\\12\catcode `\$12\catcode `\&12\catcode `\#12\catcode `\^12\catcode `\_12\catcode `\%12\relax}%
\providecommand \@@startlink[1]{}%
\providecommand \@@endlink[0]{}%
\providecommand \url  [0]{\begingroup\@sanitize@url \@url }%
\providecommand \@url [1]{\endgroup\@href {#1}{\urlprefix }}%
\providecommand \urlprefix  [0]{URL }%
\providecommand \Eprint [0]{\href }%
\providecommand \doibase [0]{https://doi.org/}%
\providecommand \selectlanguage [0]{\@gobble}%
\providecommand \bibinfo  [0]{\@secondoftwo}%
\providecommand \bibfield  [0]{\@secondoftwo}%
\providecommand \translation [1]{[#1]}%
\providecommand \BibitemOpen [0]{}%
\providecommand \bibitemStop [0]{}%
\providecommand \bibitemNoStop [0]{.\EOS\space}%
\providecommand \EOS [0]{\spacefactor3000\relax}%
\providecommand \BibitemShut  [1]{\csname bibitem#1\endcsname}%
\let\auto@bib@innerbib\@empty
\bibitem [{\citenamefont {Veneziano}(1968)}]{Veneziano:1968yb}%
  \BibitemOpen
  \bibfield  {author} {\bibinfo {author} {\bibfnamefont {G.}~\bibnamefont {Veneziano}},\ }\bibfield  {title} {\bibinfo {title} {{Construction of a crossing - symmetric, Regge behaved amplitude for linearly rising trajectories}},\ }\href {https://doi.org/10.1007/BF02824451} {\bibfield  {journal} {\bibinfo  {journal} {Nuovo Cim. A}\ }\textbf {\bibinfo {volume} {57}},\ \bibinfo {pages} {190} (\bibinfo {year} {1968})}\BibitemShut {NoStop}%
\bibitem [{\citenamefont {Paton}\ and\ \citenamefont {Chan}(1969)}]{Paton:1969je}%
  \BibitemOpen
  \bibfield  {author} {\bibinfo {author} {\bibfnamefont {J.~E.}\ \bibnamefont {Paton}}\ and\ \bibinfo {author} {\bibfnamefont {H.-M.}\ \bibnamefont {Chan}},\ }\bibfield  {title} {\bibinfo {title} {{Generalized veneziano model with isospin}},\ }\href {https://doi.org/10.1016/0550-3213(69)90038-8} {\bibfield  {journal} {\bibinfo  {journal} {Nucl. Phys. B}\ }\textbf {\bibinfo {volume} {10}},\ \bibinfo {pages} {516} (\bibinfo {year} {1969})}\BibitemShut {NoStop}%
\bibitem [{Note1()}]{Note1}%
  \BibitemOpen
  \bibinfo {note} {To see the absence of odd-spin exchange beyond the present case of four external scalars, it can be useful to adopt the spinor-helicity formalism, see e.g. \cite {Arkani-Hamed:2017jhn}.}\BibitemShut {Stop}%
\bibitem [{\citenamefont {Virasoro}(1969)}]{Virasoro:1969me}%
  \BibitemOpen
  \bibfield  {author} {\bibinfo {author} {\bibfnamefont {M.~A.}\ \bibnamefont {Virasoro}},\ }\bibfield  {title} {\bibinfo {title} {{Alternative constructions of crossing-symmetric amplitudes with regge behavior}},\ }\href {https://doi.org/10.1103/PhysRev.177.2309} {\bibfield  {journal} {\bibinfo  {journal} {Phys. Rev.}\ }\textbf {\bibinfo {volume} {177}},\ \bibinfo {pages} {2309} (\bibinfo {year} {1969})}\BibitemShut {NoStop}%
\bibitem [{\citenamefont {Shapiro}(1970)}]{Shapiro:1970gy}%
  \BibitemOpen
  \bibfield  {author} {\bibinfo {author} {\bibfnamefont {J.~A.}\ \bibnamefont {Shapiro}},\ }\bibfield  {title} {\bibinfo {title} {{Electrostatic analog for the Virasoro model}},\ }\href {https://doi.org/10.1016/0370-2693(70)90255-8} {\bibfield  {journal} {\bibinfo  {journal} {Phys. Lett. B}\ }\textbf {\bibinfo {volume} {33}},\ \bibinfo {pages} {361} (\bibinfo {year} {1970})}\BibitemShut {NoStop}%
\bibitem [{\citenamefont {Brower}\ and\ \citenamefont {Goddard}(1971)}]{Brower:1971nd}%
  \BibitemOpen
  \bibfield  {author} {\bibinfo {author} {\bibfnamefont {R.~C.}\ \bibnamefont {Brower}}\ and\ \bibinfo {author} {\bibfnamefont {P.}~\bibnamefont {Goddard}},\ }\bibfield  {title} {\bibinfo {title} {{Generalized Virasoro models}},\ }\href {https://doi.org/10.1007/BF02770367} {\bibfield  {journal} {\bibinfo  {journal} {Lett. Nuovo Cim.}\ }\textbf {\bibinfo {volume} {1S2}},\ \bibinfo {pages} {1075} (\bibinfo {year} {1971})}\BibitemShut {NoStop}%
\bibitem [{\citenamefont {Koba}\ and\ \citenamefont {Nielsen}(1969)}]{Koba:1969rw}%
  \BibitemOpen
  \bibfield  {author} {\bibinfo {author} {\bibfnamefont {Z.}~\bibnamefont {Koba}}\ and\ \bibinfo {author} {\bibfnamefont {H.~B.}\ \bibnamefont {Nielsen}},\ }\bibfield  {title} {\bibinfo {title} {{Reaction amplitude for n mesons: A Generalization of the Veneziano-Bardakci-Ruegg-Virasora model}},\ }\href {https://doi.org/10.1016/0550-3213(69)90331-9} {\bibfield  {journal} {\bibinfo  {journal} {Nucl. Phys. B}\ }\textbf {\bibinfo {volume} {10}},\ \bibinfo {pages} {633} (\bibinfo {year} {1969})}\BibitemShut {NoStop}%
\bibitem [{Note2()}]{Note2}%
  \BibitemOpen
  \bibinfo {note} {As written, the integral \protect \eqref {intgeneral} is only independent of the choice of $\protect \vec {x}_a^0$, $\protect \vec {x}_b^0$, $\protect \vec {x}_c^0$ when the external particles are tachyons with the specific mass stated in the main text. To retain the independence for other mass values, it is necessary to introduce additional powers of norms of differences to the integrand, as in the field known as $Z$-theory \cite {Carrasco:2016ldy}.}\BibitemShut {Stop}%
\bibitem [{\citenamefont {Natsuume}(1993)}]{Natsuume:1993ix}%
  \BibitemOpen
  \bibfield  {author} {\bibinfo {author} {\bibfnamefont {M.}~\bibnamefont {Natsuume}},\ }\bibfield  {title} {\bibinfo {title} {{Natural generalization of bosonic string amplitudes}},\ }\href@noop {} {\  (\bibinfo {year} {1993})},\ \Eprint {https://arxiv.org/abs/hep-th/9302131} {arXiv:hep-th/9302131} \BibitemShut {NoStop}%
\bibitem [{\citenamefont {de~Wit}\ \emph {et~al.}(1989)\citenamefont {de~Wit}, \citenamefont {Luscher},\ and\ \citenamefont {Nicolai}}]{deWit:1988xki}%
  \BibitemOpen
  \bibfield  {author} {\bibinfo {author} {\bibfnamefont {B.}~\bibnamefont {de~Wit}}, \bibinfo {author} {\bibfnamefont {M.}~\bibnamefont {Luscher}},\ and\ \bibinfo {author} {\bibfnamefont {H.}~\bibnamefont {Nicolai}},\ }\bibfield  {title} {\bibinfo {title} {{The Supermembrane Is Unstable}},\ }\href {https://doi.org/10.1016/0550-3213(89)90214-9} {\bibfield  {journal} {\bibinfo  {journal} {Nucl. Phys. B}\ }\textbf {\bibinfo {volume} {320}},\ \bibinfo {pages} {135} (\bibinfo {year} {1989})}\BibitemShut {NoStop}%
\bibitem [{\citenamefont {Green}\ and\ \citenamefont {Thorn}(1991)}]{Green:1991pa}%
  \BibitemOpen
  \bibfield  {author} {\bibinfo {author} {\bibfnamefont {M.~B.}\ \bibnamefont {Green}}\ and\ \bibinfo {author} {\bibfnamefont {C.~B.}\ \bibnamefont {Thorn}},\ }\bibfield  {title} {\bibinfo {title} {{Continuing between closed and open strings}},\ }\href {https://doi.org/10.1016/0550-3213(91)90022-P} {\bibfield  {journal} {\bibinfo  {journal} {Nucl. Phys. B}\ }\textbf {\bibinfo {volume} {367}},\ \bibinfo {pages} {462} (\bibinfo {year} {1991})}\BibitemShut {NoStop}%
\bibitem [{\citenamefont {Bjerrum-Bohr}\ and\ \citenamefont {Jepsen}(2024)}]{Bjerrum-Bohr:2024wyw}%
  \BibitemOpen
  \bibfield  {author} {\bibinfo {author} {\bibfnamefont {N.~E.~J.}\ \bibnamefont {Bjerrum-Bohr}}\ and\ \bibinfo {author} {\bibfnamefont {C.~B.}\ \bibnamefont {Jepsen}},\ }\bibfield  {title} {\bibinfo {title} {{Scattering on the worldvolume: Amplitude relations in Brower-Goddard string models}},\ }\href {https://doi.org/10.1103/PhysRevD.110.L081902} {\bibfield  {journal} {\bibinfo  {journal} {Phys. Rev. D}\ }\textbf {\bibinfo {volume} {110}},\ \bibinfo {pages} {L081902} (\bibinfo {year} {2024})},\ \Eprint {https://arxiv.org/abs/2406.10176} {arXiv:2406.10176 [hep-th]} \BibitemShut {NoStop}%
\bibitem [{\citenamefont {Gliozzi}\ \emph {et~al.}(1977)\citenamefont {Gliozzi}, \citenamefont {Scherk},\ and\ \citenamefont {Olive}}]{Gliozzi:1976qd}%
  \BibitemOpen
  \bibfield  {author} {\bibinfo {author} {\bibfnamefont {F.}~\bibnamefont {Gliozzi}}, \bibinfo {author} {\bibfnamefont {J.}~\bibnamefont {Scherk}},\ and\ \bibinfo {author} {\bibfnamefont {D.~I.}\ \bibnamefont {Olive}},\ }\bibfield  {title} {\bibinfo {title} {{Supersymmetry, Supergravity Theories and the Dual Spinor Model}},\ }\href {https://doi.org/10.1016/0550-3213(77)90206-1} {\bibfield  {journal} {\bibinfo  {journal} {Nucl. Phys. B}\ }\textbf {\bibinfo {volume} {122}},\ \bibinfo {pages} {253} (\bibinfo {year} {1977})}\BibitemShut {NoStop}%
\bibitem [{Note3()}]{Note3}%
  \BibitemOpen
  \bibinfo {note} {The convention adopted here is to strip off from the tentative superamplitude a factor of $s^2$ arising from the supermomentum-conserving delta function $\delta ^{8}(\protect \mathcal {Q})$ so that the function considered, sometimes called $f(s,t)$ in the literature, exhibits crossing symmetry: $A_s^{(2)}(s,t)=A_s^{(2)}(t,s)$.}\BibitemShut {Stop}%
\bibitem [{\citenamefont {Saha}\ and\ \citenamefont {Sinha}(2024)}]{Saha:2024qpt}%
  \BibitemOpen
  \bibfield  {author} {\bibinfo {author} {\bibfnamefont {A.~P.}\ \bibnamefont {Saha}}\ and\ \bibinfo {author} {\bibfnamefont {A.}~\bibnamefont {Sinha}},\ }\bibfield  {title} {\bibinfo {title} {{Field Theory Expansions of String Theory Amplitudes}},\ }\href {https://doi.org/10.1103/PhysRevLett.132.221601} {\bibfield  {journal} {\bibinfo  {journal} {Phys. Rev. Lett.}\ }\textbf {\bibinfo {volume} {132}},\ \bibinfo {pages} {221601} (\bibinfo {year} {2024})},\ \Eprint {https://arxiv.org/abs/2401.05733} {arXiv:2401.05733 [hep-th]} \BibitemShut {NoStop}%
\bibitem [{\citenamefont {Rosengren}(2025)}]{Rosengren:2024rpx}%
  \BibitemOpen
  \bibfield  {author} {\bibinfo {author} {\bibfnamefont {H.}~\bibnamefont {Rosengren}},\ }\bibfield  {title} {\bibinfo {title} {{String theory amplitudes and partial fractions}},\ }\href {https://doi.org/10.1007/s11139-025-01080-z} {\bibfield  {journal} {\bibinfo  {journal} {Ramanujan J.}\ }\textbf {\bibinfo {volume} {67}},\ \bibinfo {pages} {26} (\bibinfo {year} {2025})},\ \Eprint {https://arxiv.org/abs/2409.06658} {arXiv:2409.06658 [math.CA]} \BibitemShut {NoStop}%
\bibitem [{\citenamefont {Bhat}\ \emph {et~al.}(2025)\citenamefont {Bhat}, \citenamefont {Saha},\ and\ \citenamefont {Sinha}}]{Bhat2025:stringy}%
  \BibitemOpen
  \bibfield  {author} {\bibinfo {author} {\bibfnamefont {F.}~\bibnamefont {Bhat}}, \bibinfo {author} {\bibfnamefont {A.~P.}\ \bibnamefont {Saha}},\ and\ \bibinfo {author} {\bibfnamefont {A.}~\bibnamefont {Sinha}},\ }\bibfield  {title} {\bibinfo {title} {{A stringy dispersion relation for field theory}},\ }\href@noop {} {\  (\bibinfo {year} {2025})},\ \Eprint {https://arxiv.org/abs/2506.03862} {arXiv:2506.03862 [hep-th]} \BibitemShut {NoStop}%
\bibitem [{\citenamefont {Copetti}\ \emph {et~al.}(2024)\citenamefont {Copetti}, \citenamefont {Cordova},\ and\ \citenamefont {Komatsu}}]{Copetti:2024rqj}%
  \BibitemOpen
  \bibfield  {author} {\bibinfo {author} {\bibfnamefont {C.}~\bibnamefont {Copetti}}, \bibinfo {author} {\bibfnamefont {L.}~\bibnamefont {Cordova}},\ and\ \bibinfo {author} {\bibfnamefont {S.}~\bibnamefont {Komatsu}},\ }\bibfield  {title} {\bibinfo {title} {{Noninvertible Symmetries, Anomalies, and Scattering Amplitudes}},\ }\href {https://doi.org/10.1103/PhysRevLett.133.181601} {\bibfield  {journal} {\bibinfo  {journal} {Phys. Rev. Lett.}\ }\textbf {\bibinfo {volume} {133}},\ \bibinfo {pages} {181601} (\bibinfo {year} {2024})},\ \Eprint {https://arxiv.org/abs/2403.04835} {arXiv:2403.04835 [hep-th]} \BibitemShut {NoStop}%
\bibitem [{\citenamefont {Cheung}\ and\ \citenamefont {Remmen}(2023)}]{Cheung:2023adk}%
  \BibitemOpen
  \bibfield  {author} {\bibinfo {author} {\bibfnamefont {C.}~\bibnamefont {Cheung}}\ and\ \bibinfo {author} {\bibfnamefont {G.~N.}\ \bibnamefont {Remmen}},\ }\bibfield  {title} {\bibinfo {title} {{Stringy dynamics from an amplitudes bootstrap}},\ }\href {https://doi.org/10.1103/PhysRevD.108.026011} {\bibfield  {journal} {\bibinfo  {journal} {Phys. Rev. D}\ }\textbf {\bibinfo {volume} {108}},\ \bibinfo {pages} {026011} (\bibinfo {year} {2023})},\ \Eprint {https://arxiv.org/abs/2302.12263} {arXiv:2302.12263 [hep-th]} \BibitemShut {NoStop}%
\bibitem [{\citenamefont {Rigatos}(2024)}]{Rigatos:2023asb}%
  \BibitemOpen
  \bibfield  {author} {\bibinfo {author} {\bibfnamefont {K.~C.}\ \bibnamefont {Rigatos}},\ }\bibfield  {title} {\bibinfo {title} {{Positivity of the hypergeometric Veneziano amplitude}},\ }\href {https://doi.org/10.1103/PhysRevD.109.086008} {\bibfield  {journal} {\bibinfo  {journal} {Phys. Rev. D}\ }\textbf {\bibinfo {volume} {109}},\ \bibinfo {pages} {086008} (\bibinfo {year} {2024})},\ \Eprint {https://arxiv.org/abs/2310.12207} {arXiv:2310.12207 [hep-th]} \BibitemShut {NoStop}%
\bibitem [{\citenamefont {Mansfield}\ and\ \citenamefont {Spradlin}(2025)}]{Mansfield:2024wjc}%
  \BibitemOpen
  \bibfield  {author} {\bibinfo {author} {\bibfnamefont {G.}~\bibnamefont {Mansfield}}\ and\ \bibinfo {author} {\bibfnamefont {M.}~\bibnamefont {Spradlin}},\ }\bibfield  {title} {\bibinfo {title} {{On unitarity of the hypergeometric amplitude}},\ }\href {https://doi.org/10.1007/JHEP02(2025)145} {\bibfield  {journal} {\bibinfo  {journal} {JHEP}\ }\textbf {\bibinfo {volume} {02}},\ \bibinfo {pages} {145}},\ \Eprint {https://arxiv.org/abs/2409.09561} {arXiv:2409.09561 [hep-th]} \BibitemShut {NoStop}%
\bibitem [{\citenamefont {Cheung}\ \emph {et~al.}(2024)\citenamefont {Cheung}, \citenamefont {Hillman},\ and\ \citenamefont {Remmen}}]{Cheung:2024uhn}%
  \BibitemOpen
  \bibfield  {author} {\bibinfo {author} {\bibfnamefont {C.}~\bibnamefont {Cheung}}, \bibinfo {author} {\bibfnamefont {A.}~\bibnamefont {Hillman}},\ and\ \bibinfo {author} {\bibfnamefont {G.~N.}\ \bibnamefont {Remmen}},\ }\bibfield  {title} {\bibinfo {title} {{Bootstrap Principle for the Spectrum and Scattering of Strings}},\ }\href {https://doi.org/10.1103/PhysRevLett.133.251601} {\bibfield  {journal} {\bibinfo  {journal} {Phys. Rev. Lett.}\ }\textbf {\bibinfo {volume} {133}},\ \bibinfo {pages} {251601} (\bibinfo {year} {2024})},\ \Eprint {https://arxiv.org/abs/2406.02665} {arXiv:2406.02665 [hep-th]} \BibitemShut {NoStop}%
\bibitem [{\citenamefont {Cheung}\ \emph {et~al.}(2025)\citenamefont {Cheung}, \citenamefont {Hillman},\ and\ \citenamefont {Remmen}}]{Cheung:2024obl}%
  \BibitemOpen
  \bibfield  {author} {\bibinfo {author} {\bibfnamefont {C.}~\bibnamefont {Cheung}}, \bibinfo {author} {\bibfnamefont {A.}~\bibnamefont {Hillman}},\ and\ \bibinfo {author} {\bibfnamefont {G.~N.}\ \bibnamefont {Remmen}},\ }\bibfield  {title} {\bibinfo {title} {{Uniqueness criteria for the Virasoro-Shapiro amplitude}},\ }\href {https://doi.org/10.1103/PhysRevD.111.086034} {\bibfield  {journal} {\bibinfo  {journal} {Phys. Rev. D}\ }\textbf {\bibinfo {volume} {111}},\ \bibinfo {pages} {086034} (\bibinfo {year} {2025})},\ \Eprint {https://arxiv.org/abs/2408.03362} {arXiv:2408.03362 [hep-th]} \BibitemShut {NoStop}%
\bibitem [{\citenamefont {H\"aring}\ and\ \citenamefont {Zhiboedov}(2024)}]{Haring:2023zwu}%
  \BibitemOpen
  \bibfield  {author} {\bibinfo {author} {\bibfnamefont {K.}~\bibnamefont {H\"aring}}\ and\ \bibinfo {author} {\bibfnamefont {A.}~\bibnamefont {Zhiboedov}},\ }\bibfield  {title} {\bibinfo {title} {{The stringy S-matrix bootstrap: maximal spin and superpolynomial softness}},\ }\href {https://doi.org/10.1007/JHEP10(2024)075} {\bibfield  {journal} {\bibinfo  {journal} {JHEP}\ }\textbf {\bibinfo {volume} {10}},\ \bibinfo {pages} {075}},\ \Eprint {https://arxiv.org/abs/2311.13631} {arXiv:2311.13631 [hep-th]} \BibitemShut {NoStop}%
\bibitem [{Note4()}]{Note4}%
  \BibitemOpen
  \bibinfo {note} {The bound $D\leq 57$ comes from the spin-8 partial wave of the $n=10$ pole.}\BibitemShut {Stop}%
\bibitem [{\citenamefont {Siegel}(2020)}]{Siegel:2020gws}%
  \BibitemOpen
  \bibfield  {author} {\bibinfo {author} {\bibfnamefont {W.}~\bibnamefont {Siegel}},\ }\bibfield  {title} {\bibinfo {title} {{S-matrices from 4d worldvolume}},\ }\href@noop {} {\  (\bibinfo {year} {2020})},\ \Eprint {https://arxiv.org/abs/2012.12938} {arXiv:2012.12938 [hep-th]} \BibitemShut {NoStop}%
\bibitem [{\citenamefont {Arkani-Hamed}\ \emph {et~al.}(2022)\citenamefont {Arkani-Hamed}, \citenamefont {Eberhardt}, \citenamefont {Huang},\ and\ \citenamefont {Mizera}}]{Arkani-Hamed:2022gsa}%
  \BibitemOpen
  \bibfield  {author} {\bibinfo {author} {\bibfnamefont {N.}~\bibnamefont {Arkani-Hamed}}, \bibinfo {author} {\bibfnamefont {L.}~\bibnamefont {Eberhardt}}, \bibinfo {author} {\bibfnamefont {Y.-t.}\ \bibnamefont {Huang}},\ and\ \bibinfo {author} {\bibfnamefont {S.}~\bibnamefont {Mizera}},\ }\bibfield  {title} {\bibinfo {title} {{On unitarity of tree-level string amplitudes}},\ }\href {https://doi.org/10.1007/JHEP02(2022)197} {\bibfield  {journal} {\bibinfo  {journal} {JHEP}\ }\textbf {\bibinfo {volume} {02}},\ \bibinfo {pages} {197}},\ \Eprint {https://arxiv.org/abs/2201.11575} {arXiv:2201.11575 [hep-th]} \BibitemShut {NoStop}%
\bibitem [{\citenamefont {Mansfield}(2025)}]{Mansfield:2025gca}%
  \BibitemOpen
  \bibfield  {author} {\bibinfo {author} {\bibfnamefont {G.}~\bibnamefont {Mansfield}},\ }\bibfield  {title} {\bibinfo {title} {{Positivity of the Veneziano Amplitude in Ten Dimensions}},\ }\href@noop {} {\  (\bibinfo {year} {2025})},\ \Eprint {https://arxiv.org/abs/2502.20372} {arXiv:2502.20372 [hep-th]} \BibitemShut {NoStop}%
\bibitem [{\citenamefont {Eckner}\ \emph {et~al.}(2024)\citenamefont {Eckner}, \citenamefont {Figueroa},\ and\ \citenamefont {Tourkine}}]{Eckner:2024ggx}%
  \BibitemOpen
  \bibfield  {author} {\bibinfo {author} {\bibfnamefont {C.}~\bibnamefont {Eckner}}, \bibinfo {author} {\bibfnamefont {F.}~\bibnamefont {Figueroa}},\ and\ \bibinfo {author} {\bibfnamefont {P.}~\bibnamefont {Tourkine}},\ }\bibfield  {title} {\bibinfo {title} {{The Regge bootstrap, from linear to non-linear trajectories}},\ }\href@noop {} {\  (\bibinfo {year} {2024})},\ \Eprint {https://arxiv.org/abs/2401.08736} {arXiv:2401.08736 [hep-th]} \BibitemShut {NoStop}%
\bibitem [{\citenamefont {Caron-Huot}\ and\ \citenamefont {Van~Duong}(2021)}]{Caron-Huot:2020cmc}%
  \BibitemOpen
  \bibfield  {author} {\bibinfo {author} {\bibfnamefont {S.}~\bibnamefont {Caron-Huot}}\ and\ \bibinfo {author} {\bibfnamefont {V.}~\bibnamefont {Van~Duong}},\ }\bibfield  {title} {\bibinfo {title} {{Extremal Effective Field Theories}},\ }\href {https://doi.org/10.1007/JHEP05(2021)280} {\bibfield  {journal} {\bibinfo  {journal} {JHEP}\ }\textbf {\bibinfo {volume} {05}},\ \bibinfo {pages} {280}},\ \Eprint {https://arxiv.org/abs/2011.02957} {arXiv:2011.02957 [hep-th]} \BibitemShut {NoStop}%
\bibitem [{\citenamefont {Arkani-Hamed}\ \emph {et~al.}(2021{\natexlab{a}})\citenamefont {Arkani-Hamed}, \citenamefont {Huang},\ and\ \citenamefont {Huang}}]{Arkani-Hamed:2020blm}%
  \BibitemOpen
  \bibfield  {author} {\bibinfo {author} {\bibfnamefont {N.}~\bibnamefont {Arkani-Hamed}}, \bibinfo {author} {\bibfnamefont {T.-C.}\ \bibnamefont {Huang}},\ and\ \bibinfo {author} {\bibfnamefont {Y.-t.}\ \bibnamefont {Huang}},\ }\bibfield  {title} {\bibinfo {title} {{The EFT-Hedron}},\ }\href {https://doi.org/10.1007/JHEP05(2021)259} {\bibfield  {journal} {\bibinfo  {journal} {JHEP}\ }\textbf {\bibinfo {volume} {05}},\ \bibinfo {pages} {259}},\ \Eprint {https://arxiv.org/abs/2012.15849} {arXiv:2012.15849 [hep-th]} \BibitemShut {NoStop}%
\bibitem [{\citenamefont {Berman}\ \emph {et~al.}(2024)\citenamefont {Berman}, \citenamefont {Elvang},\ and\ \citenamefont {Herderschee}}]{Berman:2023jys}%
  \BibitemOpen
  \bibfield  {author} {\bibinfo {author} {\bibfnamefont {J.}~\bibnamefont {Berman}}, \bibinfo {author} {\bibfnamefont {H.}~\bibnamefont {Elvang}},\ and\ \bibinfo {author} {\bibfnamefont {A.}~\bibnamefont {Herderschee}},\ }\bibfield  {title} {\bibinfo {title} {{Flattening of the EFT-hedron: supersymmetric positivity bounds and the search for string theory}},\ }\href {https://doi.org/10.1007/JHEP03(2024)021} {\bibfield  {journal} {\bibinfo  {journal} {JHEP}\ }\textbf {\bibinfo {volume} {03}},\ \bibinfo {pages} {021}},\ \Eprint {https://arxiv.org/abs/2310.10729} {arXiv:2310.10729 [hep-th]} \BibitemShut {NoStop}%
\bibitem [{\citenamefont {Chiang}\ \emph {et~al.}(2024)\citenamefont {Chiang}, \citenamefont {Huang},\ and\ \citenamefont {Weng}}]{Chiang:2023quf}%
  \BibitemOpen
  \bibfield  {author} {\bibinfo {author} {\bibfnamefont {L.-Y.}\ \bibnamefont {Chiang}}, \bibinfo {author} {\bibfnamefont {Y.-t.}\ \bibnamefont {Huang}},\ and\ \bibinfo {author} {\bibfnamefont {H.-C.}\ \bibnamefont {Weng}},\ }\bibfield  {title} {\bibinfo {title} {{Bootstrapping string theory EFT}},\ }\href {https://doi.org/10.1007/JHEP05(2024)289} {\bibfield  {journal} {\bibinfo  {journal} {JHEP}\ }\textbf {\bibinfo {volume} {05}},\ \bibinfo {pages} {289}},\ \Eprint {https://arxiv.org/abs/2310.10710} {arXiv:2310.10710 [hep-th]} \BibitemShut {NoStop}%
\bibitem [{\citenamefont {Berman}\ and\ \citenamefont {Elvang}(2024)}]{Berman:2024wyt}%
  \BibitemOpen
  \bibfield  {author} {\bibinfo {author} {\bibfnamefont {J.}~\bibnamefont {Berman}}\ and\ \bibinfo {author} {\bibfnamefont {H.}~\bibnamefont {Elvang}},\ }\bibfield  {title} {\bibinfo {title} {{Corners and islands in the S-matrix bootstrap of the open superstring}},\ }\href {https://doi.org/10.1007/JHEP09(2024)076} {\bibfield  {journal} {\bibinfo  {journal} {JHEP}\ }\textbf {\bibinfo {volume} {09}},\ \bibinfo {pages} {076}},\ \Eprint {https://arxiv.org/abs/2406.03543} {arXiv:2406.03543 [hep-th]} \BibitemShut {NoStop}%
\bibitem [{\citenamefont {Boulanger}\ \emph {et~al.}(2000)\citenamefont {Boulanger}, \citenamefont {Damour}, \citenamefont {Gualtieri},\ and\ \citenamefont {Henneaux}}]{Boulanger:2000bp}%
  \BibitemOpen
  \bibfield  {author} {\bibinfo {author} {\bibfnamefont {N.}~\bibnamefont {Boulanger}}, \bibinfo {author} {\bibfnamefont {T.}~\bibnamefont {Damour}}, \bibinfo {author} {\bibfnamefont {L.}~\bibnamefont {Gualtieri}},\ and\ \bibinfo {author} {\bibfnamefont {M.}~\bibnamefont {Henneaux}},\ }\bibfield  {title} {\bibinfo {title} {{No consistent cross interactions for a collection of massless spin-2 fields}},\ }\href@noop {} {\bibfield  {journal} {\bibinfo  {journal} {Ann. U. Craiova Phys.}\ }\textbf {\bibinfo {volume} {10}},\ \bibinfo {pages} {94} (\bibinfo {year} {2000})},\ \Eprint {https://arxiv.org/abs/hep-th/0009109} {arXiv:hep-th/0009109} \BibitemShut {NoStop}%
\bibitem [{\citenamefont {Boulanger}\ \emph {et~al.}(2001)\citenamefont {Boulanger}, \citenamefont {Damour}, \citenamefont {Gualtieri},\ and\ \citenamefont {Henneaux}}]{Boulanger:2000rq}%
  \BibitemOpen
  \bibfield  {author} {\bibinfo {author} {\bibfnamefont {N.}~\bibnamefont {Boulanger}}, \bibinfo {author} {\bibfnamefont {T.}~\bibnamefont {Damour}}, \bibinfo {author} {\bibfnamefont {L.}~\bibnamefont {Gualtieri}},\ and\ \bibinfo {author} {\bibfnamefont {M.}~\bibnamefont {Henneaux}},\ }\bibfield  {title} {\bibinfo {title} {{Inconsistency of interacting, multigraviton theories}},\ }\href {https://doi.org/10.1016/S0550-3213(00)00718-5} {\bibfield  {journal} {\bibinfo  {journal} {Nucl. Phys. B}\ }\textbf {\bibinfo {volume} {597}},\ \bibinfo {pages} {127} (\bibinfo {year} {2001})},\ \Eprint {https://arxiv.org/abs/hep-th/0007220} {arXiv:hep-th/0007220} \BibitemShut {NoStop}%
\bibitem [{\citenamefont {Jepsen}(2023)}]{Jepsen:2023sia}%
  \BibitemOpen
  \bibfield  {author} {\bibinfo {author} {\bibfnamefont {C.~B.}\ \bibnamefont {Jepsen}},\ }\bibfield  {title} {\bibinfo {title} {{Cutting the Coon amplitude}},\ }\href {https://doi.org/10.1007/JHEP06(2023)114} {\bibfield  {journal} {\bibinfo  {journal} {JHEP}\ }\textbf {\bibinfo {volume} {06}},\ \bibinfo {pages} {114}},\ \Eprint {https://arxiv.org/abs/2303.02149} {arXiv:2303.02149 [hep-th]} \BibitemShut {NoStop}%
\bibitem [{\citenamefont {Witten}(2015)}]{Witten:2013pra}%
  \BibitemOpen
  \bibfield  {author} {\bibinfo {author} {\bibfnamefont {E.}~\bibnamefont {Witten}},\ }\bibfield  {title} {\bibinfo {title} {{The Feynman $i \epsilon$ in String Theory}},\ }\href {https://doi.org/10.1007/JHEP04(2015)055} {\bibfield  {journal} {\bibinfo  {journal} {JHEP}\ }\textbf {\bibinfo {volume} {04}},\ \bibinfo {pages} {055}},\ \Eprint {https://arxiv.org/abs/1307.5124} {arXiv:1307.5124 [hep-th]} \BibitemShut {NoStop}%
\bibitem [{\citenamefont {Eberhardt}\ and\ \citenamefont {Mizera}(2023)}]{Eberhardt:2023xck}%
  \BibitemOpen
  \bibfield  {author} {\bibinfo {author} {\bibfnamefont {L.}~\bibnamefont {Eberhardt}}\ and\ \bibinfo {author} {\bibfnamefont {S.}~\bibnamefont {Mizera}},\ }\bibfield  {title} {\bibinfo {title} {{Evaluating one-loop string amplitudes}},\ }\href {https://doi.org/10.21468/SciPostPhys.15.3.119} {\bibfield  {journal} {\bibinfo  {journal} {SciPost Phys.}\ }\textbf {\bibinfo {volume} {15}},\ \bibinfo {pages} {119} (\bibinfo {year} {2023})},\ \Eprint {https://arxiv.org/abs/2302.12733} {arXiv:2302.12733 [hep-th]} \BibitemShut {NoStop}%
\bibitem [{\citenamefont {Eberhardt}\ and\ \citenamefont {Mizera}(2024)}]{Eberhardt:2024twy}%
  \BibitemOpen
  \bibfield  {author} {\bibinfo {author} {\bibfnamefont {L.}~\bibnamefont {Eberhardt}}\ and\ \bibinfo {author} {\bibfnamefont {S.}~\bibnamefont {Mizera}},\ }\bibfield  {title} {\bibinfo {title} {{Lorentzian contours for tree-level string amplitudes}},\ }\href {https://doi.org/10.21468/SciPostPhys.17.3.078} {\bibfield  {journal} {\bibinfo  {journal} {SciPost Phys.}\ }\textbf {\bibinfo {volume} {17}},\ \bibinfo {pages} {078} (\bibinfo {year} {2024})},\ \Eprint {https://arxiv.org/abs/2403.07051} {arXiv:2403.07051 [hep-th]} \BibitemShut {NoStop}%
\bibitem [{\citenamefont {Manschot}\ and\ \citenamefont {Wang}(2024)}]{Manschot:2024prc}%
  \BibitemOpen
  \bibfield  {author} {\bibinfo {author} {\bibfnamefont {J.}~\bibnamefont {Manschot}}\ and\ \bibinfo {author} {\bibfnamefont {Z.-Z.}\ \bibnamefont {Wang}},\ }\bibfield  {title} {\bibinfo {title} {{The $i\varepsilon$-Prescription for String Amplitudes and Regularized Modular Integrals}},\ }\href@noop {} {\  (\bibinfo {year} {2024})},\ \Eprint {https://arxiv.org/abs/2411.02517} {arXiv:2411.02517 [hep-th]} \BibitemShut {NoStop}%
\bibitem [{\citenamefont {Kapranov}(2021)}]{Kapranov:2021nfy}%
  \BibitemOpen
  \bibfield  {author} {\bibinfo {author} {\bibfnamefont {M.}~\bibnamefont {Kapranov}},\ }\bibfield  {title} {\bibinfo {title} {{Conformal maps in higher dimensions and derived geometry}},\ }\href@noop {} {\  (\bibinfo {year} {2021})},\ \Eprint {https://arxiv.org/abs/2102.11507} {arXiv:2102.11507 [math.AG]} \BibitemShut {NoStop}%
\bibitem [{\citenamefont {Arkani-Hamed}\ \emph {et~al.}(2024)\citenamefont {Arkani-Hamed}, \citenamefont {Cheung}, \citenamefont {Figueiredo},\ and\ \citenamefont {Remmen}}]{Arkani-Hamed:2023jwn}%
  \BibitemOpen
  \bibfield  {author} {\bibinfo {author} {\bibfnamefont {N.}~\bibnamefont {Arkani-Hamed}}, \bibinfo {author} {\bibfnamefont {C.}~\bibnamefont {Cheung}}, \bibinfo {author} {\bibfnamefont {C.}~\bibnamefont {Figueiredo}},\ and\ \bibinfo {author} {\bibfnamefont {G.~N.}\ \bibnamefont {Remmen}},\ }\bibfield  {title} {\bibinfo {title} {{Multiparticle Factorization and the Rigidity of String Theory}},\ }\href {https://doi.org/10.1103/PhysRevLett.132.091601} {\bibfield  {journal} {\bibinfo  {journal} {Phys. Rev. Lett.}\ }\textbf {\bibinfo {volume} {132}},\ \bibinfo {pages} {091601} (\bibinfo {year} {2024})},\ \Eprint {https://arxiv.org/abs/2312.07652} {arXiv:2312.07652 [hep-th]} \BibitemShut {NoStop}%
\bibitem [{\citenamefont {Cheung}\ and\ \citenamefont {Remmen}(2025)}]{Cheung:2025krg}%
  \BibitemOpen
  \bibfield  {author} {\bibinfo {author} {\bibfnamefont {C.}~\bibnamefont {Cheung}}\ and\ \bibinfo {author} {\bibfnamefont {G.~N.}\ \bibnamefont {Remmen}},\ }\bibfield  {title} {\bibinfo {title} {{Multipositivity Bounds}},\ }\href@noop {} {\  (\bibinfo {year} {2025})},\ \Eprint {https://arxiv.org/abs/2505.05553} {arXiv:2505.05553 [hep-th]} \BibitemShut {NoStop}%
\bibitem [{\citenamefont {Arkani-Hamed}\ \emph {et~al.}(2023)\citenamefont {Arkani-Hamed}, \citenamefont {Frost}, \citenamefont {Salvatori}, \citenamefont {Plamondon},\ and\ \citenamefont {Thomas}}]{Arkani-Hamed:2023lbd}%
  \BibitemOpen
  \bibfield  {author} {\bibinfo {author} {\bibfnamefont {N.}~\bibnamefont {Arkani-Hamed}}, \bibinfo {author} {\bibfnamefont {H.}~\bibnamefont {Frost}}, \bibinfo {author} {\bibfnamefont {G.}~\bibnamefont {Salvatori}}, \bibinfo {author} {\bibfnamefont {P.-G.}\ \bibnamefont {Plamondon}},\ and\ \bibinfo {author} {\bibfnamefont {H.}~\bibnamefont {Thomas}},\ }\bibfield  {title} {\bibinfo {title} {{All Loop Scattering As A Counting Problem}},\ }\href@noop {} {\  (\bibinfo {year} {2023})},\ \Eprint {https://arxiv.org/abs/2309.15913} {arXiv:2309.15913 [hep-th]} \BibitemShut {NoStop}%
\bibitem [{\citenamefont {Cao}\ \emph {et~al.}(2025)\citenamefont {Cao}, \citenamefont {Dong}, \citenamefont {He},\ and\ \citenamefont {Zhu}}]{Cao:2025lzv}%
  \BibitemOpen
  \bibfield  {author} {\bibinfo {author} {\bibfnamefont {Q.}~\bibnamefont {Cao}}, \bibinfo {author} {\bibfnamefont {J.}~\bibnamefont {Dong}}, \bibinfo {author} {\bibfnamefont {S.}~\bibnamefont {He}},\ and\ \bibinfo {author} {\bibfnamefont {F.}~\bibnamefont {Zhu}},\ }\bibfield  {title} {\bibinfo {title} {{Superstring amplitudes meet surfaceology}},\ }\href@noop {} {\  (\bibinfo {year} {2025})},\ \Eprint {https://arxiv.org/abs/2504.21676} {arXiv:2504.21676 [hep-th]} \BibitemShut {NoStop}%
\bibitem [{\citenamefont {Arkani-Hamed}\ \emph {et~al.}(2021{\natexlab{b}})\citenamefont {Arkani-Hamed}, \citenamefont {Huang},\ and\ \citenamefont {Huang}}]{Arkani-Hamed:2017jhn}%
  \BibitemOpen
  \bibfield  {author} {\bibinfo {author} {\bibfnamefont {N.}~\bibnamefont {Arkani-Hamed}}, \bibinfo {author} {\bibfnamefont {T.-C.}\ \bibnamefont {Huang}},\ and\ \bibinfo {author} {\bibfnamefont {Y.-t.}\ \bibnamefont {Huang}},\ }\bibfield  {title} {\bibinfo {title} {{Scattering amplitudes for all masses and spins}},\ }\href {https://doi.org/10.1007/JHEP11(2021)070} {\bibfield  {journal} {\bibinfo  {journal} {JHEP}\ }\textbf {\bibinfo {volume} {11}},\ \bibinfo {pages} {070}},\ \Eprint {https://arxiv.org/abs/1709.04891} {arXiv:1709.04891 [hep-th]} \BibitemShut {NoStop}%
\bibitem [{\citenamefont {Carrasco}\ \emph {et~al.}(2017)\citenamefont {Carrasco}, \citenamefont {Mafra},\ and\ \citenamefont {Schlotterer}}]{Carrasco:2016ldy}%
  \BibitemOpen
  \bibfield  {author} {\bibinfo {author} {\bibfnamefont {J.~J.~M.}\ \bibnamefont {Carrasco}}, \bibinfo {author} {\bibfnamefont {C.~R.}\ \bibnamefont {Mafra}},\ and\ \bibinfo {author} {\bibfnamefont {O.}~\bibnamefont {Schlotterer}},\ }\bibfield  {title} {\bibinfo {title} {{Abelian Z-theory: NLSM amplitudes and $\alpha$'-corrections from the open string}},\ }\href {https://doi.org/10.1007/JHEP06(2017)093} {\bibfield  {journal} {\bibinfo  {journal} {JHEP}\ }\textbf {\bibinfo {volume} {06}},\ \bibinfo {pages} {093}},\ \Eprint {https://arxiv.org/abs/1608.02569} {arXiv:1608.02569 [hep-th]} \BibitemShut {NoStop}%
\end{thebibliography}%

\end{document}